\documentclass[useAMS,usenatbib]{coin}

\def\bSig\mathbf{\Sigma}

\usepackage[figuresright]{rotating}

\usepackage{hyperref}

\usepackage{amssymb}
\usepackage{mathtools}
\usepackage{mathpazo}
\usepackage{float}
\usepackage{booktabs}
\usepackage{bm}
\usepackage{enumitem}
\usepackage{listings}
\usepackage{multicol}
\usepackage{multirow}
\usepackage{hhline}

\usepackage{amssymb}
\usepackage{amsmath}

\usepackage{url}
\usepackage{latexsym}
\usepackage{booktabs}

\usepackage{graphicx}

\usepackage{amssymb}
\usepackage{amsmath}

\usepackage{booktabs}

\usepackage{hhline}

\usepackage{tablefootnote}
\usepackage{algorithmicx}
\usepackage{algorithm}
\usepackage{algpseudocode}

\usepackage{siunitx}
\usepackage{subfig}

\usepackage{bm}

\usepackage{capt-of}

\usepackage{stackengine}

\usepackage{enumitem}

\usepackage{todonotes}

\usepackage{listings}
\usepackage{tabularx}

\usepackage{array}
\newcolumntype{N}{>{\centering\arraybackslash}m{.5in}}
\newcolumntype{G}{>{\centering\arraybackslash}m{2in}}
\newcolumntype{L}[1]{>{\raggedright\arraybackslash}p{#1}}

\RequirePackage[most]{tcolorbox}

\usepackage{colortbl}
\definecolor{table:lefttext}{RGB}{64,64,64}

\DeclareFixedFont{\ttb}{T1}{txtt}{bx}{n}{12} 
\DeclareFixedFont{\ttm}{T1}{txtt}{m}{n}{12}  

\usepackage{color}
\definecolor{deepblue}{rgb}{0,0,0.5}
\definecolor{deepred}{rgb}{0.6,0,0}
\definecolor{deepgreen}{rgb}{0,0.5,0}

\definecolor{cblue}{RGB}{68,122,187}

\usepackage[format=plain,
labelfont={it},
textfont=it]{caption}

\usepackage{listings}

\newcommand\pythonstyle{\lstset{
		language=Python,
		basicstyle=\ttm,
		otherkeywords={self},             
		keywordstyle=\color{deepblue},
		emph={MyClass,__init__},          
		emphstyle=\ttb\color{deepred},    
		stringstyle=\color{deepgreen},
		frame=tb,                         
		showstringspaces=false            %
	}}

	\lstnewenvironment{python}[1][]
	{
		\pythonstyle
		\lstset{#1}
	}
	{}
	
	\newcommand\pythoninline[1]{{\pythonstyle\lstinline!#1!}}

	\newcommand{\swat}{\textsc{Swat}}
	
	\let\customMakeTitle=\maketitle
	\renewcommand{\maketitle}{\begingroup\let\footnote=\thanks \customMakeTitle\endgroup}

	\title{\swat{}: A System for Detecting Salient Wikipedia Entities in Texts}

	\author{{\sc Marco Ponza, Paolo Ferragina, Francesco Piccinno}\\
		{\it University of Pisa}\\
		\texttt{\{firstname.lastname\}@di.unipi.it}
	}

	\hypersetup{pdfcreator={},pdfproducer={}}

\begin{document}

		\pagerange{\pageref{firstpage}--\pageref{lastpage}} \pubyear{2019}
		
		\label{firstpage}

		\begin{abstract}
			We study the problem of entity salience by proposing the design and implementation of \textsc{Swat}, a system that identifies the salient Wikipedia entities occurring in an input document. \textsc{Swat} consists of several modules that are able to detect and classify on-the-fly Wikipedia entities as salient or not, based on a large number of syntactic, semantic and latent features properly extracted via a supervised process which has been trained over millions of examples drawn from the New York Times corpus. The validation process is performed through a large experimental assessment, eventually showing that \textsc{Swat} improves known solutions over all publicly available datasets. We release  \textsc{Swat} via an API that we describe~and~comment~in~the~paper in order to ease its use in other software.
		\end{abstract}

		
		\maketitle

		\footnotetext{The present paper is an \textit{extended} version of the one published in the {\em Proceedings of the 22nd International Conference on Natural Language \& Information Systems.} \citep{ponza2017document}.}	
		\footnotetext{This work has been published at \textit{Computational Intelligence, Wiley-Blackwell Publishing (2019)} and it is available at the address \href{https://doi.org/10.1111/coin.12216}{https://doi.org/10.1111/coin.12216}.}
		
		\section{Introduction}
		
		Detecting \textit{salient} information in documents, such as sentences~\citep{mihalcea2004textrank},  open facts~\citep{ponza2018facts}, keywords~\citep{bruza1996study,paranjpe2009learning,hasan2014automatic}, or Wikipedia entities~\citep{dunietz2014new,trani2016sel}, has become a fundamental task on which different Information Retrieval (IR) and Natural Language Processing (NLP) tools hinge upon to improve their performance. Contextual ads-matching~\citep{radlinski2008optimizing}, document similarity~\citep{ni2016semantic}, web search ranking~\citep{gamon2013identifying,schuhmacher2015ranking}, and news suggestion~\citep{fetahu2015automated} are just a few examples of typical research domains on which the salient information extracted from natural language texts is consumed.

		In this paper we propose a new system called \swat{} ({\textbf{S}alient \textbf{W}ikipedia \textbf{A}nnota\-tion of \textbf{T}ext), which constitutes the state-of-the-art in detecting salient Wikipedia entities occurring in an input text. The software architecture of \swat{} relies on a pipeline organized in three main modules: {Document Enrichment}, {Feature Generation} and {Entity Salience Classification}. Given an input document, the  {Document Enrichment} module annotates it with proper syntactic, semantic and latent information that are automatically extracted through the deployment of four software components: (i) {CoreNLP} \citep{manning2014stanford} --- the most well-known NLP framework to analyze the grammatical structure of sentences --- is used to extract the morphological information coming from the dependency trees built over the sentences of the input document; (ii) \textsc{Wat}~\citep{piccinno2014tagme} --- one of the best publicly available entity linkers~\citep{usbeck2015gerbil} --- is used to annotate the text with proper Wikipedia entities and to build an entity graph for weighting the importance of these entities and their semantic relationships; (iii) {TextRank}~\citep{mihalcea2004textrank} --- the popular document summarizer --- is used to return a keyphrase score for each sentence of the input document; and (iv) {Word2Vec} --- the continuous vector space representation of words and entities captured via deep neural networks --- is used to enrich the entity graph of point (ii) with distributional latent signals. Subsequently, the {Feature Generation} module dispatches the enriched information generated from the first stage to a number of other software components in order to map each entity into its proper vector of features, which significantly expands the ones investigated in previous papers~\citep{dunietz2014new,trani2016sel}. Finally, these feature vectors are fed to the {Entity Salience Classification} module that leads to discriminate entities into salient and non-salient.
			
			The validation of our system is performed through a large experimental assessment executed over two datasets, known as New York Times and Wikinews. \textsc{Swat} is compared against two systems that constitute the state-of-the-art in this setting, namely \textsc{Cmu-Google} \citep{dunietz2014new} and \textsc{Sel} \citep{trani2016sel}. This experimental study shows that \textsc{Swat} raises the best-known performance in terms of F1 up to $3.4\%$ (absolute) over \textsc{Cmu-Google} system and up to $6.3\%$ (absolute) over \textsc{Sel} system in either of the two experimented datasets. These F1-results are complemented with a throughout discussion about the impact of each feature  onto the overall performance of our system and on how the position of salient entities does influence the efficacy of their detection. In this latter setting, we show that the improvement of \textsc{Swat} with respect to \textsc{Cmu-Google} over the largest dataset New York Times may get up to $14\%$ in micro-F1.
			
			\medskip\noindent Summarizing, the main contributions of the paper are the following ones:
			
			\begin{itemize}

				\item We design and implement \textsc{Swat}, an effective entity salience system that detects the salient entities of a document via the design and use of a novel and  rich set of syntactic (e.g., sentences' ranking and dependency trees), latent (i.e., word and entity embeddings), and semantic (i.e., computed via a new graph representation of entities and several centrality measures) features. Despite the use of word and entity embeddings is not new in IR, we are the first (to the best of our knowledge) to investigate its effectiveness on the entity salience task with a proper engineering of features based on two different latent representations of entities.

				\item We are the first ones to offer an extensive experimental comparison among all known entity salience systems (i.e., \textsc{Swat}, \textsc{Sel} and \textsc{Cmu-Google}, plus several other baselines) over the available datasets: New York Times~\citep{dunietz2014new} and Wikinews~\citep{trani2016sel}.

				\item The experiments show that \textsc{Swat} consistently improves the F1 performance of \textsc{Cmu-Google} and \textsc{Sel} over these two datasets by achieving, respectively, an improvement of about $12.2\%$ (absolute) and $6.3\%$ (absolute). 
				
				\item These figures are accompanied by a thoughtful analysis of \textsc{Swat}'s features, efficiency and errors, thus showing that all of its components are crucial to achieve its improved performance both in F1 and time efficiency. 
				
				\item In order to encourage the development of other research built upon entity salience tools, we release \textsc{Swat} as a public API\footnote{\label{footenote:swatapi} \href{https://sobigdata.d4science.org/web/tagme/swat-api}{https://sobigdata.d4science.org/web/tagme/swat-api}}, which actually implements the full entity-linking-and-salience pipeline thus ease its plugging into other software.
			\end{itemize}
			
			\noindent The paper is organized as follows. Section \ref{sec:related} discusses the problem of the detection of salient information in texts by presenting known solutions and their limitations. Section \ref{sec:swat} describes the design principles at the core of \swat{} by detailing its three main  modules and posing particular attention on the sophisticated and novel feature extraction process. Section \ref{sec:exp} digs into the experimental comparison between \swat{} and the current state-of-the-art systems --- i.e., \textsc{Cmu-Google} \citep{dunietz2014new} and \textsc{Sel} \citep{trani2016sel} --- over the New York Times and Wikinews datasets. The experimental figures show a coherent and significant improvement of \swat{} with respect to these systems over both datasets. The subsequent Section \ref{sec:discussion} extends the previous experimental analysis with a discussion on four engineering and algorithmic aspects pertaining to the design of \swat{}: (i) the impact that the features have on the quality of its entity salient predictions, (ii) its efficiency in terms of constituting modules and used features, (ii) the impact of the training-set size onto its generalization ability and, finally, (iv) a thoughtful error analysis that will highlight the deficiencies of the known datasets. Taking inspiration from the previous detailed discussion, Section \ref{sec:conclusions} introduces several interesting research directions which would be worth to be investigated in the near future because could lead to further improvements on the solution to the entity salience task.

			\section{Related Work}
			\label{sec:related}
			
			Classical approaches for detecting salient information in documents are known under the umbrella topic of  \textit{keyphrase extraction}~\citep{hasan2014automatic}. These systems identify keyphrases through the  \textit{lexical} elements of the input text, such as words labeled with specific  POS tags~\citep{mihalcea2004textrank,liu2010automatic,gamon2013identifying}, n-grams~\citep{turney2000learning} or words that belong to a fixed dictionary of terms~\citep{paranjpe2009learning}. The salient keyphrases are then selected from these lexical elements via supervised or unsupervised machine learning~\citep{paranjpe2009learning,gamon2013identifying}. Unfortunately, key\-phrase extraction systems commonly incur in several limitations which have been properly highlighted in the previous literature~\citep{hasan2014automatic}: (i) their interpretation is left to the reader (i.e., \textit{ interpretation errors}); (ii) words that appear frequently in the input text often induce  the selection of non-salient keyphrases (i.e., \textit{ over-generation errors}); (iii) infrequent keyphrases go undetected (i.e., \textit{ infrequency errors}); and (iv) by working at a pure lexical level the keyphrase-based systems are unable to detect the semantic equivalence between two keyphrases (i.e., \textit{ redundancy errors}).
			
			Given these limitations, some researchers tried to introduce some ``semantics'' into the salient representation of a document by taking advantage of the recent progress in the design of entity linking systems (see \citep{shen2015entity} and references therein). The key idea underlying these approaches consists of identifying in the input text meaningful sequences of terms and link them to \textit{unambiguous} entities drawn from a Knowledge Base (KB), such as Wikipedia, DBpedia~\citep{bizer2009dbpedia}, Freebase \citep{bollacker2008freebase}, Wikidata~\citep{vrandevcic2014wikidata}, YAGO \citep{suchanek2007yago}, or BabelNet \citep{navigli2012babelnet}. Since these entities occur as nodes in a graph, new and more sophisticated methods have been designed in order to empower classic approaches and thus enabling a number of significant improvements among different domains, such as microblog enrichment and analysis \citep{ferragina2015analyzing,liu2013entity,meij2012adding}, text classification and clustering \citep{scaiella2012topical,VitaleFS12}, KB construction \citep{niu2012deepdive,bovi2015large,nguyen2017query} and query understanding \citep{blanco2015fast,hasibi2017entity,cornolti2019}. 
			
			On the other hand, assigning a proper \textit{salient} label to Wikipedia entities is still in its infancy and, indeed, only two approaches are known: namely, the \textsc{Cmu-Google} \citep{dunietz2014new} system and the \textsc{Sel}~\citep{trani2016sel} system. The first one uses a proprietary entity linker to extract entities from the input text and a binary classifier based on very few and simple features to distinguish between salient and non-salient entities. Dunietz and Gillick (2014) have shown that their system significantly outperforms a simple baseline via some experiments executed over the large and well-known New York Times dataset. Unfortunately, the software deploys proprietary modules that make it publicly unavailable. In the end, authors concluded that: \textit{ ``There is likely significant room for improvement, [$\ldots$]. Perhaps features more directly linked to Wikipedia, as in related work on keyword extraction, can provide more focused background information''.}  
			
			Following this intuition, Trani \textit{et al.} (2017) proposed the second known approach, called \textsc{Sel}, that hinges on a supervised two-step algorithm comprehensively addressing both entity linking and entity salience. The first step is based on a classifier aimed at identifying a set of candidate entities that are mentioned in the document, thus maximizing the precision without hindering the recall; the second step is based on a regression model that aims at  scoring the candidate entities. Unfortunately \textsc{Sel} was compared only against \textit{pure} entity linkers --- such as \textsc{TagMe} \citep{FerraginaS12} --- which were not designed for the entity salience task, the system is yet publicly unavailable and, furthermore, its experimental figures were confined to a new dataset (i.e., Wikinews), which is much smaller than NYT, and thus missed a comparison against the~\textsc{Cmu-Google} system. 
			
			As a result, the two entity salience systems above are not publicly available and their experimental figures are incomparable. In the present paper, we continue the study of the entity salience problem by introducing a novel system, that we call \textsc{Swat}, whose main goal is to efficiently and efficaciously address these open issues through the improvement of the state-of-the-art.
			
				\vspace{0.4cm}	
				\subsection{Background}	
				
				In this work, we investigate the use of a number of known NLP/IR tools that we apply for the design of our novel system \textsc{Swat}. Accordingly, we devote this section to recall the main concepts and general ideas  on which these algorithmic tools have been designed. Specifically, we start from the area of entity linking and comment on how their algorithms are commonly implemented. Then, after describing several tools for the general-purpose NLP analysis, we move the attention on the topic of automatic text summarization, thus presenting several similarities with respect to the entity salience task. Finally, we conclude this section with a description of the approaches for learning the latent representation of entities.
				
				\medskip\noindent\textbf{Entity Linking.} Entity linkers are tools that aim at providing a mapping between a text to the entities of a KB. More precisely, they address the task of identifying short sequences of terms (called spots or mentions) in the input text and then annotate them with unambiguous entities which belong to the KB at hand. Literature offers different publicly available solutions \citep{piccinno2014tagme,hoffart2011robust,ganea2016probabilistic,zwicklbauer2016robust}, as well as proper benchmarking platforms \citep{cornolti2013framework,usbeck2015gerbil} for their fair and extensive evaluation.
				
				Most of entity linkers usually work in a pipeline of two main stages. In the first stage, mentions are detected through the use of common NLP toolkits --- e.g., {CoreNLP} \citep{manning2014stanford} --- and then, in the second stage, they are disambiguated by associating every mention to exactly one single entity. Disambiguation algorithms commonly model this task as an optimization problem whose objective function is designed to maximize the coherence among the annotated entities. For efficiency reasons, entity linkers propose approximated solutions for this optimization problem, which is actually solved by means of several heuristics. For example, \textsc{TagMe}~\citep{scaiella2012topical} and \textsc{Wat}~\citep{piccinno2014tagme} restrict the annotation process only on those entities which receive the highest number of votes with respect to a voting scheme executed over the candidates generated for each mention; \textsc{Pboh}~\citep{ganea2016probabilistic} uses the Loopy Belief Propagation scheme for inferring entity co-occurrence probabilities, while \textsc{DoSeR}~\citep{zwicklbauer2016robust} iteratively disambiguates the entities that show stable coherence scores computed via Personalized PageRank.

				\smallskip\noindent\textbf{General-purpose NLP Analysis.} Literature currently offers a number solution for the general-purpose analysis of natural language texts.  {GATE} \citep{cunningham2002gate}, {NLTK} \citep{bird2004nltk}, {UIMA} and {DKPro} \citep{ferrucci2004uima,gurevych2007darmstadt}, {CoreNLP}~\citep{manning2014stanford}, spaCy~\citep{honnibal2017spacy}, and AllenNLP~\citep{gardner2018allennlp}, are just few examples of open-source software that provide a set of state-of-the-art NLP analyzers for natural language texts, such as POS tagging, NER, dependency annotation, and coreference resolution. These tools are very popular, especially because a variegate set of researchers, coming from different communities, have been able to built a  number of downstream applications  within a myriad of different contexts \citep{hirschberg2015advances}, such as Open Information Extraction \citep{gashteovski2017minie}, Sentiment Analysis \citep{socher2013recursive} and Question Answering \citep{chen2017reading}, just to mention a few .

			    \smallskip\noindent\textbf{Automatic Text Summarization.} This is research area concerns the extraction of a summary from an input text \citep{gambhir2017recent}. More precisely, text summarizers aim at the identifying relevant and topical information from an input text and condense them into a small set of textual elements: a text which is shorter than the original one but that still preserves the salient elements that it conveys. Summaries are clearly fundamental from different points of views. They can enable fast and accurate search of documents  from large text collections \citep{hasan2014automatic} as well as they can help a reader to immediately identify the relevant topics of the original document. The domain of document summarization can actually be clustered among different dimensions~\citep{gambhir2017recent}, such as the summary's {objective} (i.e., generic vs query-focused), the number of documents  to summarize (i.e., single vs multi-document) and the summarizer's approach (i.e., extractive vs abstractive). Accordingly, entity salience can be seen as a subfield of extractive, generic and single-document summarization where the summary is expressed as a set of salient entities.
				
				\smallskip\noindent\textbf{Entity Embeddings.} Word embeddings \citep{mikolov2013distributed} is a recent technique which aims at mapping words into low dimensional numerical vectors. This latent representation has been recently extended to learn the embeddings of entities through two main algorithmic approaches: Entity2Vec~\citep{ni2016semantic} and DeepWalk~\citep{perozzi2014deepwalk}.  The former approach aims at learning the embedding of entities by exploiting the textual content of Wikipedia articles in order to declare similar two entities when they frequently co-occur within similar textual contexts. Complementary, the second approach aims at learning the embedding of entities by exploiting the graph structure of Wikipedia (defined with its hyperlinks) in order to declare similar two entities when a random surfer frequently walks over similar paths rooted on a specific focus node.

			\section{\textsc{Swat}: A Novel Entity Salience System}\label{swat}
			\label{sec:swat}

			In this section, we describe our system \textsc{Swat}, which aims at identifying the salient Wikipedia entities of an input document through a pipeline of three main modules: {Document Enrichment}, {Feature Generation} and {Entity Salience Classification}. A graphical representation of \swat{} is provided by Figure~\ref{fig:swatarch}.

			\begin{description}
				\setlength\itemsep{0.2cm}
				
				\item[Document Enrichment.] The first module aims at enriching the input document $d$ with a set of semantic, morphological, syntactic and latent information. Specifically, this module is organized in four main components:
				
				\begin{enumerate}
					\setlength\itemsep{0.1cm}
					\item {CoreNLP}~\citep{manning2014stanford} is the component in charge of enriching the document with proper morphological NLP annotations. Specifically, it tokenizes the input document $d$, assigns the POS tags to the tokens, generates the dependency relations, identifies noun phrases, and finally produces the coreference chains.
					
					\item {TextRank}~\citep{mihalcea2004textrank} is a component that works by taking as input the sentences tokenized by {CoreNLP} and by rank them via a random walk over a complete graph in which nodes are sentences and the weights of the edges are computed as a function of the normalized number of common tokens between the connected sentences.
					
					\item \textsc{Wat}~\citep{piccinno2014tagme} is the component that aims to enrich $d$ with a set of semantic annotations $(m,e)$, where $m$ is a sequence of words (i.e., \textit{ mentions}, provided by {CoreNLP} as noun phrase and $e$ is an entity (i.e., Wikipedia page). Specifically, \textsc{Wat} disambiguates every mention $m$ by assigns to each mention an entity provided with two main scores: (i) \textit{commonness}, which represents the probability that $m$ is disambiguated by $e$; (ii) \textit{coherence} (denoted by $\rho$), which represents the semantic coherence between the annotation and its textual~context.
					
					Subsequently, this component generates an \textit{entity graph} in which nodes are the annotated entities and edges are weighted with the relatedness between the edge-connected entities (Jaccard Relatednesses in Figure \ref{fig:swatarch}).

					\item {Word2Vec} is the component that aims to enrich the input document with latent information. More precisely, it takes the entities annotated by \textsc{Wat} and map them into their proper continuous vector representations learned via neural networks~\citep{mikolov2013distributed}. These latent representations are further used to compute the cosine similarities between all entities that have been annotated in the document $d$ by \textsc{Wat}. Technically speaking, the {Word2Vec} component is constituted by two sub-components that respectively deploy two different kinds of { latent entity representations}: Entity2Vec~\citep{ni2016semantic} and DeepWalk~\citep{perozzi2014deepwalk} (more details are provided in Section~\ref{sec:more-on-feature-generation}), respectively.
					
				\end{enumerate}
				
				\item[Feature Generation.] The second module deploys the data generated by Document Enrichment in order to compute a rich set of features for each entity $e$.  Specifically, four main components are deployed (i.e., \texttt{Standard}, \texttt{Syntactic}, \texttt{Semantic} and \texttt{Word2Vec} in the Feature Generation module) in order to map each $e$ into its proper vector of features. A more detailed description of these components, as well as the algorithms implemented to generate the features for each entity, is provided below.

					\begin{figure}[t!]
						\centering
						\includegraphics[width=1.\textwidth]{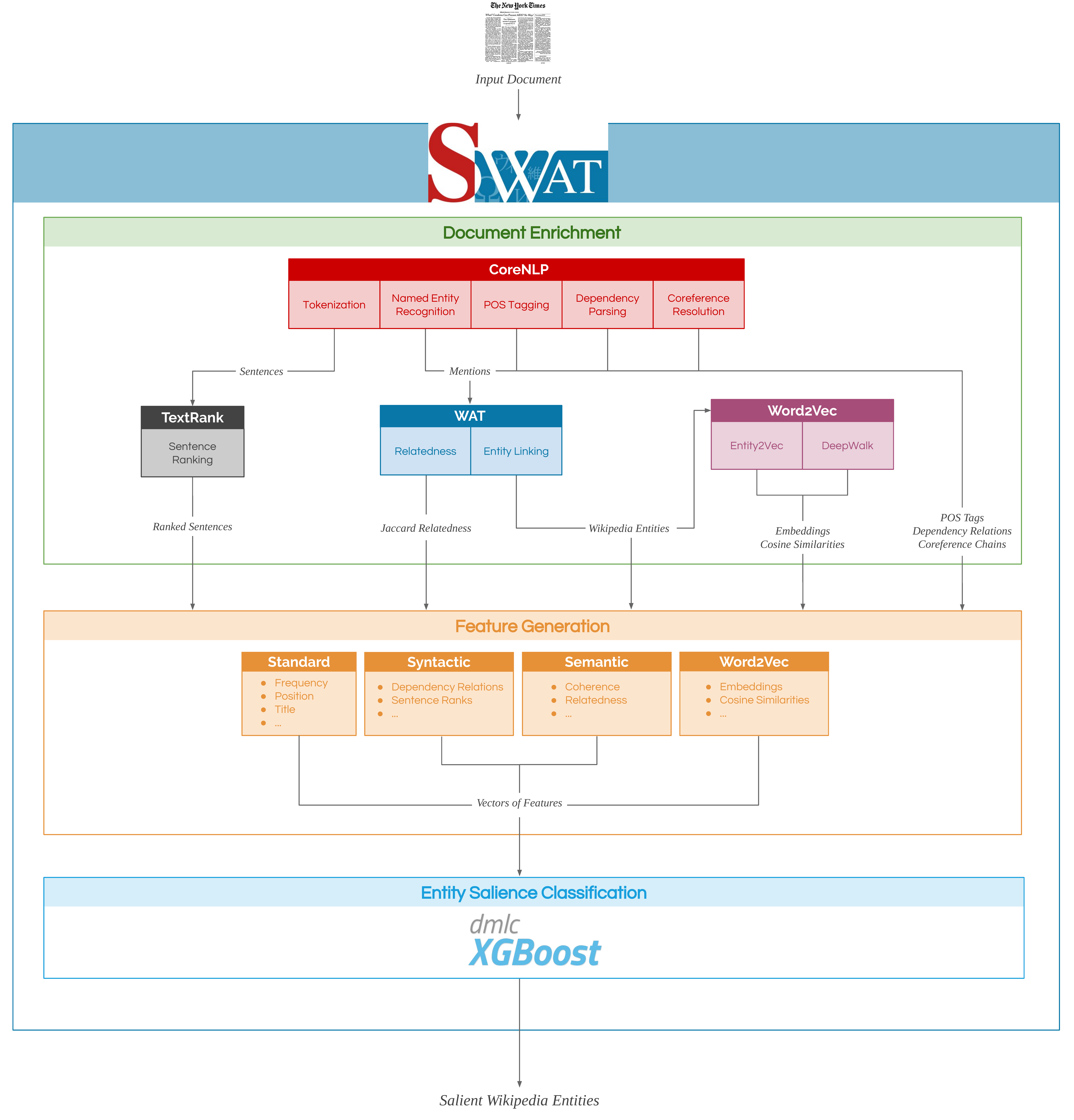}
						\vspace{0.5cm}
						\caption{Three-module architecture of \textsc{Swat}.\label{fig:swatarch}}
					\end{figure}

				\item[Entity Salience Classification.] The goal of the last module is to classify entities into their class (i.e., salient vs non-salient) given the entity features computed by the previous module. We implement this classification step through the deployment of the efficient and highly scalable {eXtreme Gradient Boosting} software library ~\citep{chen2016xgboost} (XGBoost Classifier in Figure \ref{fig:swatarch}) which is trained and tested as detailed in Section \ref{sec:exp}.
				
			\end{description}

			\begin{table}[t]

				\centering
				\caption{Standard features adopted by \swat{}.}
				\label{table:standard-features}
				
				\resizebox{1.0\textwidth}{!}{
					\begin{tabular}{p{4.5cm}p{10cm}l}
						\toprule
						\textbf{Name} & \textbf{Description} & \textbf{Component}   \\ \hline

						\addlinespace $ef (e, d), idf (e), ef \mbox{-}idf (e, d)$  & Entity frequency (number of times \textsc{Wat} annotates $e$ in $d$), inverse document frequency for $e$ and their product. & \texttt{Standard} \\
						
						\addlinespace$position\mbox{-}stats_{\{s, t\}} (e, d)$ & Minimum, maximum, arithmetic mean, median, standard deviation and harmonic mean of sentence- (resp. token-) positions of $e$ in $d$. & \texttt{Standard} \\
						
						\addlinespace	$mention\mbox{-}title(e, d)$ & Presence of a $mention$ of $e$ in the title of $d$.  & \texttt{Standard}  \\  
						\addlinespace	$entity\mbox{-}title(e, d)$   & Presence of $e$ in the title of $d$.   & \texttt{Standard}\\  
						\addlinespace	$is\mbox{-}upper(e, d)$       & True if one of the mentions of $e$ appear in $d$ in uppercase, false otherwise.   & \texttt{Standard}\\  \bottomrule
					\end{tabular}
				}
				
				\vspace{1cm}
				
				\centering
				\caption{Features introduced by the \textsc{Cmu-Google} system \citep{dunietz2014new} and adopted by \swat{}.}
				\label{table:cmu-google-features}
				
				\resizebox{1.0\textwidth}{!}{
					\begin{tabular}{p{4.5cm}p{10cm}l}
						\toprule
						\textbf{Name} & \textbf{Description} & \textbf{Component}   \\ \hline
						\addlinespace$1st \mbox{-} loc (e, d)$    & Index of the sentence in which the first mention of $e$ appears in $d$.& \texttt{Standard}\\ 
						\addlinespace$head \mbox{-} count (e, d)$ & Frequency of head word of entity $e$ in the document $d$.& \texttt{Syntactic}\\ 
						\addlinespace$mentions (e, d)$            & Sum between entity frequency and co-referenced frequency of $e$ in $d$.& \texttt{Syntactic}\\ 
						\addlinespace$headline(e, d)$             & POS tag of each word of $e$ that appears in at least one mention and also in the headline 	of $d$. & \texttt{Syntactic} \\
						\addlinespace$head \mbox{-} lex (e, d)$   & Lower-cased head word of the first mention of $e$ in $d$. & \texttt{Syntactic} \\ 
						\addlinespace$google\mbox{-}centrality(e, d)$           & PageRank score of $e$ on the entity graph generated from $d$, where weights are the co-occurrence probability of two entities, computed on the training set. & \texttt{Standard} \\
						\midrule \\
					\end{tabular}
				}
				
				\vspace{0.5cm}

			\end{table}

			\begin{table*}[htbp]
				\vspace{1cm}
				
				\centering
				\caption{Novel features introduced by \swat{}.}
				\label{table:novel-features}
				
				\resizebox{0.9\textwidth}{!}{
					\begin{tabular}{p{4.5cm}p{10.2cm}l}
						\toprule
						\textbf{Name} & \textbf{Description} & \textbf{Component}   \\ \hline
						\addlinespace $spread_{\{s, t\}} (e, d)$                        & Difference between the max and min sentence- (resp. token-) positions of $e$ in $d$. & \texttt{Standard} \\
						\addlinespace	$bucketed\mbox{-}freq_{\{s, t\}} (e, d)$                         & Vector of bucketed frequencies through sentence- (resp. token-) positions \mbox{of $e$ in $d$.} & \texttt{Standard} \\
						\addlinespace	$textrank \mbox{-} stats (e, d)$ & Minimum, maximum, arithmetic mean, median, standard deviation and harmonic mean of {TextRank} scores of sentences where $e$ appears in $d$. & \texttt{Syntactic} \\
						\addlinespace	$dep \mbox{-} freq (e, d)$                        & Frequency of $e$ in $d$ when it appears as dependent of the dependency \mbox{relation $dep$}.  & \texttt{Syntactic} \\
						\addlinespace	$dep \mbox{-}bucketed\mbox{-} freq_ {\{s, t\}} (e, d)$                        & Vector of bucketed frequencies through sentence- (resp. token-) positions of $e$ in $d$ limited to the mentions where $e$ appears as dependent with relation	$dep$. & \texttt{Syntactic} \\
						\addlinespace	$dep \mbox{-} position\mbox{-}stats_{\{s, t\}} (e, d)$              & Minimum, maximum, arithmetic mean, median, standard deviation and harmonic mean  of sentence- (resp. token-) positions of $e$ in $d$, where only the mentions where $e$ appears as dependent of a dependency relation $dep$ are considered. & \texttt{Syntactic} \\
						\addlinespace	$dep\mbox{-}textrank\mbox{-}stats (e, d)$                        & Minimum, maximum, arithmetic mean, median, standard deviation and harmonic mean  of {TextRank} scores where only the sentences where $e$ appears as dependent of a dependency relation $dep$ are taken into account. & \texttt{Syntactic} \\
						\addlinespace	$comm \mbox{-} stats (e, d)$  & Minimum, maximum, arithmetic mean, median, standard deviation and harmonic mean of the $commonness$ values of $e$ in $d$ computed by \textsc{Wat}. & \texttt{Semantic} \\
						\addlinespace	$\rho \mbox{-} stats(e, d)$   & Minimum, maximum, arithmetic mean, median, standard deviation and harmonic mean of the $\rho$-score values of $e$ in $d$ computed by \textsc{Wat}. & \texttt{Semantic} \\
						\addlinespace	$rel \mbox{-} stats (e, d)$                       & Minimum, maximum, arithmetic mean, median, standard deviation and harmonic mean  of the relatedness scores between $e$ and all other entities \mbox{annotated in $d$}. & \texttt{Semantic} \\
						\addlinespace	$rel\mbox{-}bucketed\mbox{-}stats_{\{s, t\}} (e, d)$ & Minimum, maximum, arithmetic mean, median, standard deviation and harmonic mean of the relatedness scores between $e$ and all other entities present in $d$, bucketed over document positions (both at sentence- and token-level). & \texttt{Semantic} \\
						\addlinespace	$rel\mbox{-}centrality(e, d)$ & Degree, PageRank, Betweenness, Katz, HITS, Closeness, and Harmonic scores of $e$ computed on the entity graph of~$d$.  & \texttt{Semantic} \\
						\addlinespace	$wiki \mbox{-} id (e)$       & Wikipedia identifier of $e$, normalized via feature hashing. & \texttt{Semantic} \\
						
						\addlinespace	$w2v(e)$ & Entity2Vec and DeepWalk embedding vectors of $e$. & \texttt{Word2Vec} \\
						\addlinespace	$w2v\mbox{-}stats(e, d)$ & Minimum, maximum, arithmetic mean, median, standard deviation and harmonic mean of the cosine similarity between the Entity2Vec and DeepWalk embeddings of $e$ and the ones of the other entities annotated in the title and headline of $d$. & \texttt{Word2Vec} \\
						\addlinespace	$w2v\mbox{-}cos\mbox{-}title(e, d)$ & Cosine-similarity between the Entity2Vec and DeepWalk embeddings of $e$ and the average of the corresponding embeddings of the words present in the title of~$d$. & \texttt{Word2Vec}  \\ 
						\addlinespace	$w2v\mbox{-}cos\mbox{-}headline(e, d)$ & Cosine-similarity between the Entity2Vec and DeepWalk embeddings of $e$ and the average of the corresponding embeddings of the words present in the headline of~$d$. & \texttt{Word2Vec}  \\ 
						
						\bottomrule
						
					\end{tabular}
				}
				
				\vspace{0.5cm}

			\end{table*}

			\renewcommand{\arraystretch}{1}

			\subsection{More on Feature Generation}\label{sec:more-on-feature-generation}
			
			Despite the use of the third module is pretty standard, the first and second modules are more involved and constitute the main novel part of our system \swat{}. Hence, the rest of this section is devoted to detail the first two modules which generate the features for each entity that has been annotated in the input document $d$ --- called \texttt{Standard}, \texttt{Syntactic} , \texttt{Semantic} and \texttt{Word2Vec} --- to be used in the third and last entity salience classification module. In order to facilitate the reading and understanding of the large number of features deployed by \swat{}, we report all of them in Tables \ref{table:standard-features}, \ref{table:cmu-google-features} and \ref{table:novel-features} which respectively group the features by novelty and by the software component which is in charge of their implementation (rightmost column in each table).
			
			In the text below we now comment only the new features introduced by \swat{}. For each of them, we first report their technical description and then we introduce a specific paragraph in which we detail the motivations and the phenomena that our new engineered features aim to capture.

			\bigskip\noindent\textbf{Position-based Features.} These features deploy the distribution within document $d$ of the entities occurrences in order to predict their salience score. Furthermore, all position features of an entity $e$ within the document $d$ are computed by \textsc{Swat} in terms of \textit{tokens} or \textit{sentences}. For token-level features (indicated with the subscript $t$) it is considered the index of the first token for each mention of $e$, normalized by the number of tokens of $d$; whereas for sentence-level features (indicated with the subscript $s$) it is considered the index of the sentences where the entity $e$ is annotated, normalized by the number of sentences of $d$.
			These features naturally improve the $1st\mbox{-}loc(e,d)$ feature introduced by~\cite{dunietz2014new} thus making more robust \textsc{Swat} with respect to the positional distribution of salient entities.

				\smallskip \textit{Captured Phenomena.} Through the proposal of this set of features (expressed both at sentence- and token-level) we aim at capturing finer-grain positions of the annotated entities than the one previously proposed by  \cite{dunietz2014new} with $1st\mbox{-}loc(e,d)$. More precisely, $1st\mbox{-}loc(e,d)$ is calculated by modeling the sentence index $i$ as an array of size 10 where all elements are 0 except the one at index $log( 10 \cdot ( i + 1 ) )$ --- see \citep{dunietz2014new} for details. This normalization technique  has several disadvantages: (i) it lacks in distinguishing the position among entities annotated within the same sentence, (ii) it assigns the same value to entities that are annotated in different sentences, and (iii) it totally ignores the full length of the text. A graphical example that shows these limitations and how token- and sentence-level features can solve this problem is shown in Figure~\ref{fig:example:features-position}. As we can see, \textsf{Barack Obama} and \textsf{Hilary Clinton} appear in the same sentence but at different positions, thus that $1st\mbox{-}loc$ and $position\mbox{-}min_{s}$ get the same values, but $position\mbox{-}min_t$ allows to differentiate them. Furthermore, \textsf{Iraq} and \textsf{New Hampshire}, which have been respectively annotated in the 4th and 7th sentences, achieve the same $1st\mbox{-}loc$ values but $position\mbox{-}min_{s}$ and $position\mbox{-}min_t$ get different values.
				
				\begin{figure}[t!]
					\centering
					\includegraphics[scale=0.25]{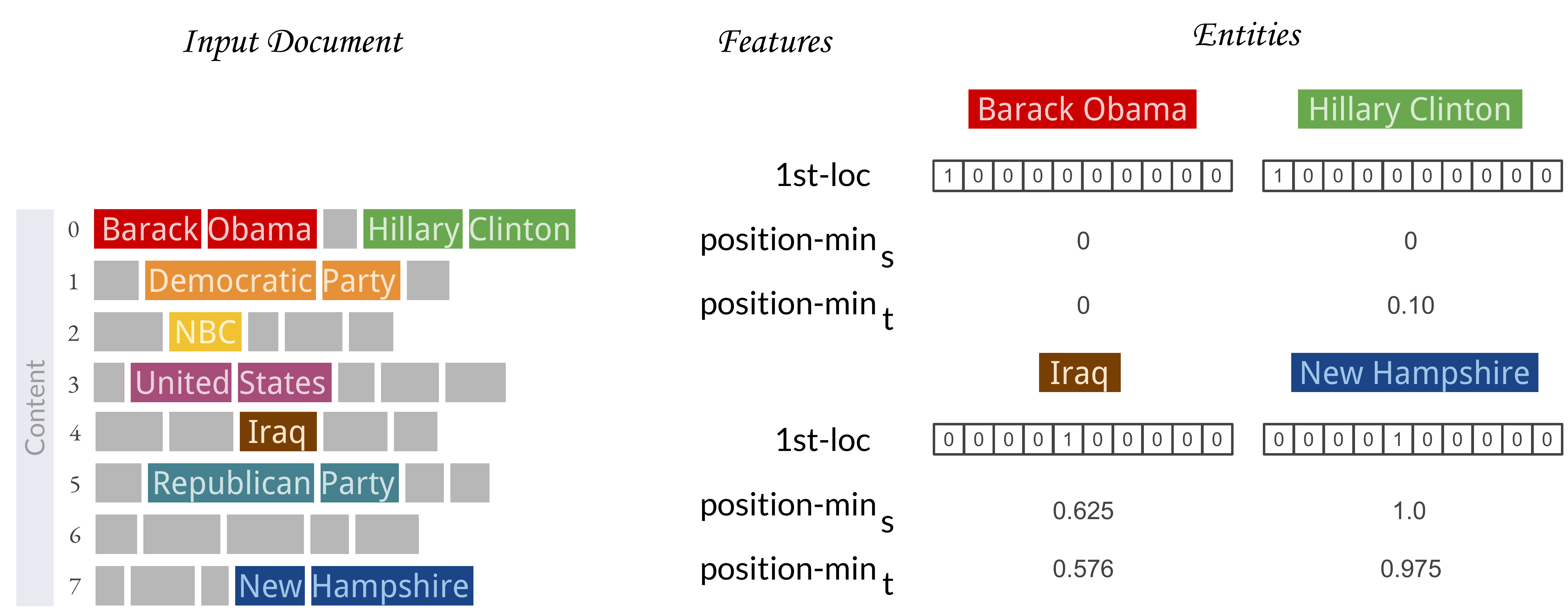}
					\vspace{0.2cm}
					\caption{
						Example that shows where $1st\mbox{-}loc$ feature fails in a proper differentiation among entities' positions. Numbers on the left of the document are the sentence indices, rectangles represent tokens in the input document and entities are tinted with different colors.}
					
					\label{fig:example:features-position}

				\end{figure}

				\begin{figure}

					\centering
					\includegraphics[scale=0.32]{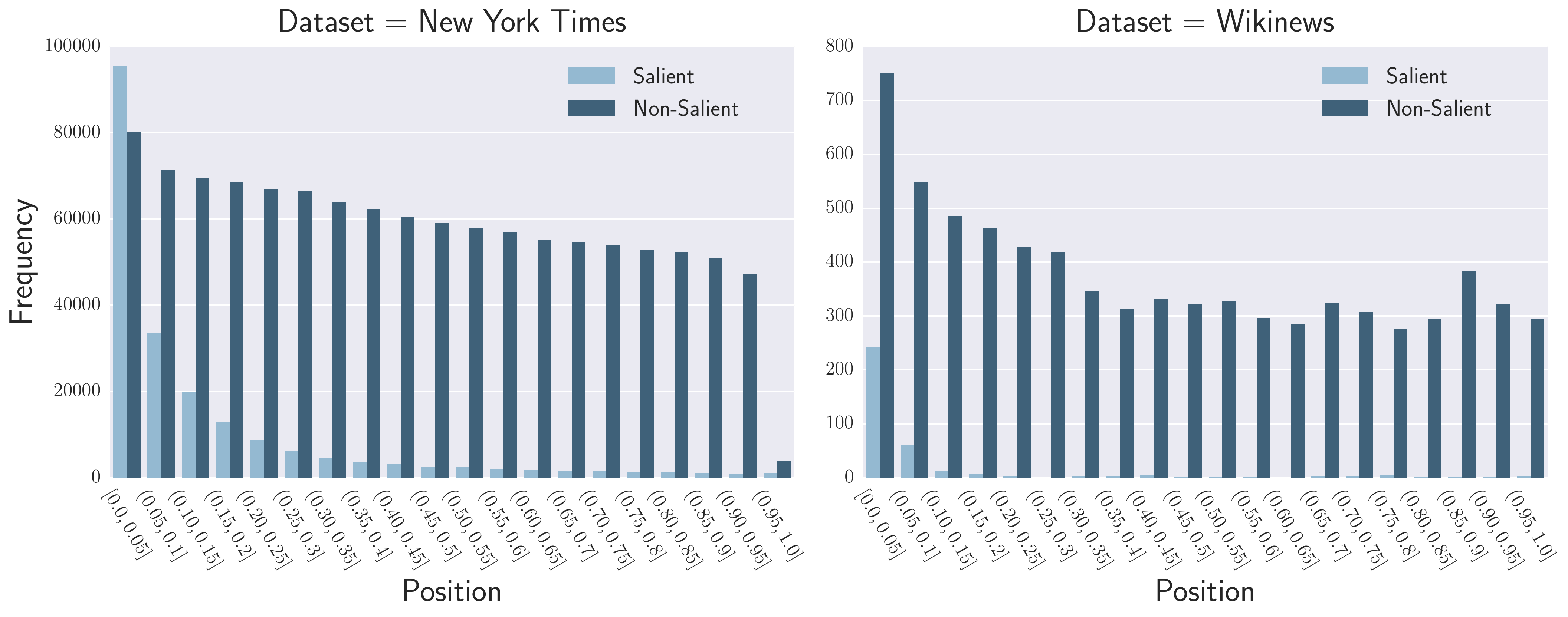}
					\vspace{0.1cm}
					\caption{The histograms plot the frequency distribution of salient versus non-salient entities according to their \textit{first} positions in the documents over NYT (left) and Wikinews datasets (right).} \label{fig:nyt-wn-pos}
					
					\vspace{0.2cm}
					
					\centering
					\includegraphics[scale=0.32]{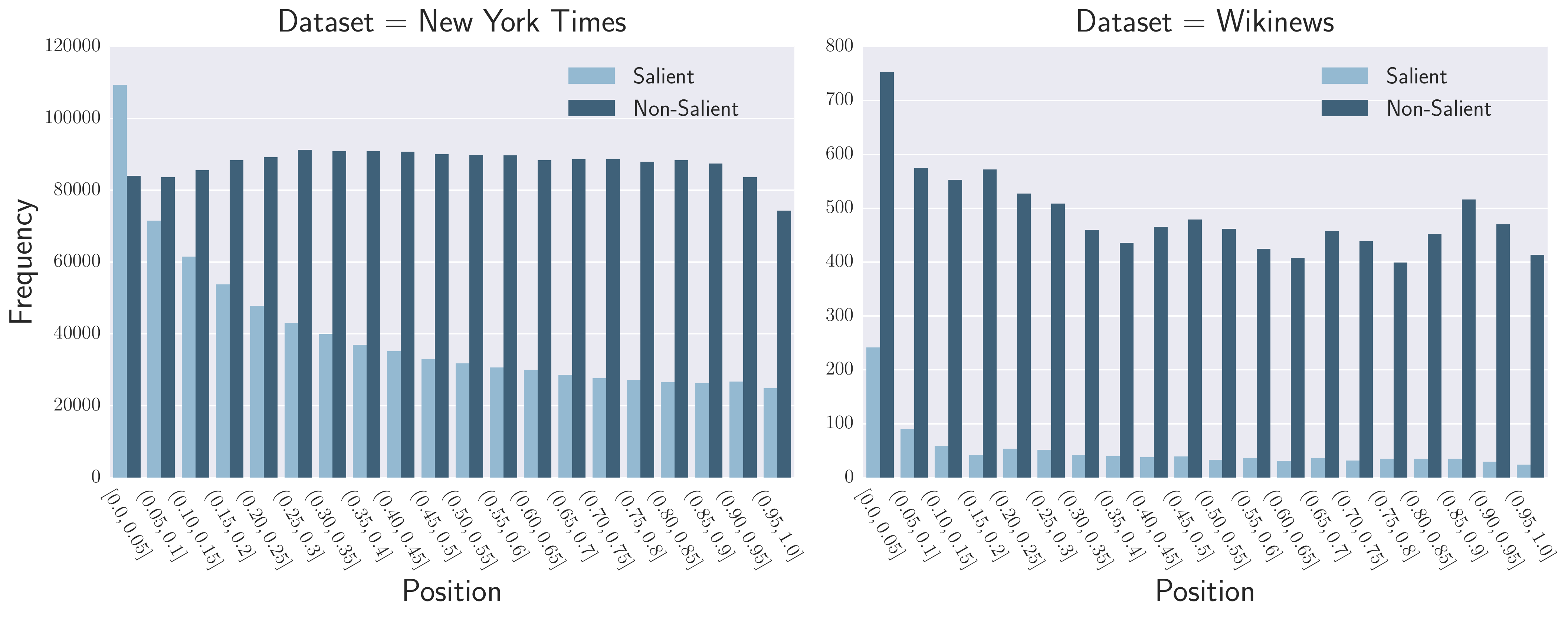}
					\vspace{0.1cm}
					\caption{The histograms plot the frequency distribution of salient versus non-salient entities according to \textit{all} their occurrences in the documents over NYT (left) and Wikinews datasets (right).} \label{fig:all-nyt-wn-pos}

				\end{figure}

				One more issue that afflicts the $1st\mbox{-}loc$ feature consists in the fact that it models only the \textit{first} position of an entity $e$ in $d$, thus failing in capturing the (possibly meaningful) distribution of that entity in the input document. Figure~\ref{fig:nyt-wn-pos} and \ref{fig:all-nyt-wn-pos} show the distribution of salient versus non-salient entities among two different datasets. As we can see, salient entities present a common pattern, with a frequency that is very high at the beginning and smoothly decreases among the rest of the document. Accordingly, we decided to investigate the computation of two specific features that should model this phenomenon: $bucketed\mbox{-}freq$ captures the distribution of an entity in the input text, and $spread$ computes the difference between the position of the first and the position of the last mention of an entity in the input text.

			\bigskip\noindent\textbf{Summarization-based Features.} These features exploit the sc\-o\-re that summarization algorithms assign to sentences that contain salient information and thus possibly contain salient entities. Accordingly, \textsc{Swat} computes, for each entity $e$, several statistical measures derived from the scores assigned by {TextRank}~\citep{mihalcea2004textrank} to the sentences where a mention of $e$ occurs.

				\smallskip\textit{Captured Phenomena.} These features aim at capturing the syntactical centrality of a sentence in a document and postulate that \textit{``salient entities are contained in sentences which are central for the input document''}. This ``centrality issue'' is a signal commonly used by popular state-of-the-art text summarizers~\citep{mihalcea2004textrank}. We implement this idea by defining a set of features that assigns high scores to entities which occur in sentences highly rated by {TextRank}.

			\bigskip \noindent\textbf{Linguistic-based Features.} These features exploit the grammatical structure, namely the dependency trees, of sentences where the entities occur. Unlike~\citep{dunietz2014new}, where dependency trees are used to extract only the head of a mention, \textsc{Swat} combines frequency, position and summarization information with several dependency relations generated by the {CoreNLP}'s dependency parser.

				\smallskip\textit{Captured Phenomena.} Through these features we aim at modeling the morphological associations (i.e., dependency relations) among the entity's mentions in a text. More precisely, a number of mentions of salient entities in the benchmarked datasets frequently have tokens which are dependent of  preposition-in, adjective modifier, possessive, noun compound modifier and subject dependency relations. Accordingly, we design features that compute position, frequency and sentence scores by prior filtering only the mention of entities whose tokens appear as dependent of the main dependency relations mentioned above (i.e., $dep\mbox{-}*$ features).

			\newpage
			\bigskip\noindent\textbf{Annotation-based Features.} This set of features computes several statistics upon $commonness$ and $\rho$ scores which have been assigned to each annotation $(m, e)$ by the entity linker \textsc{Wat}. These two scores capture two different aspects of a given annotated entity $(m, e)$: $commonness$ provides a sort of common-sense probability that $m$ can be disambiguated with $e$, whereas $\rho$ quantifies the quality of the annotation in terms of coherence between $e$ and its context of occurrence in the input document $d$.

				\smallskip\textit{Captured Phenomena.} Despite entity linkers currently reach very good performance on different datasets \citep{usbeck2015gerbil}, they can also incur into several errors by annotating a mention $m$ with a misleading entity $e$. In the entity salience problem, a wrongly annotated entity can introduce some noise in the entity salience pipeline, with a worst-case scenario where the misleading entity is eventually classified as salient. To limit the impact of entities wrongly detected by \textsc{Wat}, we decided to extract several other features based on the $commonness$ and $\rho$ scores (resp. $comm\mbox{-}$ and $\rho\mbox{-}stats$ features) with the intuition that these scores should increase the robustness of our entity salient classifier.

			\bigskip\noindent\textbf{{Word2Vec}-based Features.} This set of features aims at modeling the annotated entities and their relationships by means of proper embeddings generated via deep neural networks. Specifically, \textsc{Swat} deploys the well-known CBOW and Skip-gram models~\citep{mikolov2013distributed} here applied to entities by means of two algorithms:
			
			\begin{enumerate}
				\item Entity2Vec~\citep{ni2016semantic} is an extension of the original {Word2Vec} that computes a unique embedding for both entities and words extracted from the textual descriptions of the Wikipedia pages.
				
				\item DeepWalk~\citep{perozzi2014deepwalk} is another variation of the original {Word2Vec} that  computes an embedding for nodes of a graph, which is here the Wikipedia graph. 
			\end{enumerate}
			
			\noindent\textsc{Swat} uses as features the continuous vectors derived from Entity2Vec and DeepWalk, plus several other statistics computed over their cosine similarity measure.

				\smallskip\textit{Captured Phenomena.} The features built on top of the {Word2Vec} component aim at capturing those kinds of latent signals that cannot be explicitly detected from the syntactic and morphological features proposed before. For example, the latent relationships between the title of the input document and the candidate entities should help our system to correctly detect the correct entities, especially when they are salient and do not appear at the beginning. A graphical example is reported in Figure~\ref{fig:features-title}. As we can see, the title contains information that can actually help \textsc{Swat} to correctly classify \textsf{Silvester Stallone} and \textsf{Muhammed Ali} as salient.

			\begin{figure}[t]
				\centering
				\includegraphics[scale=0.25]{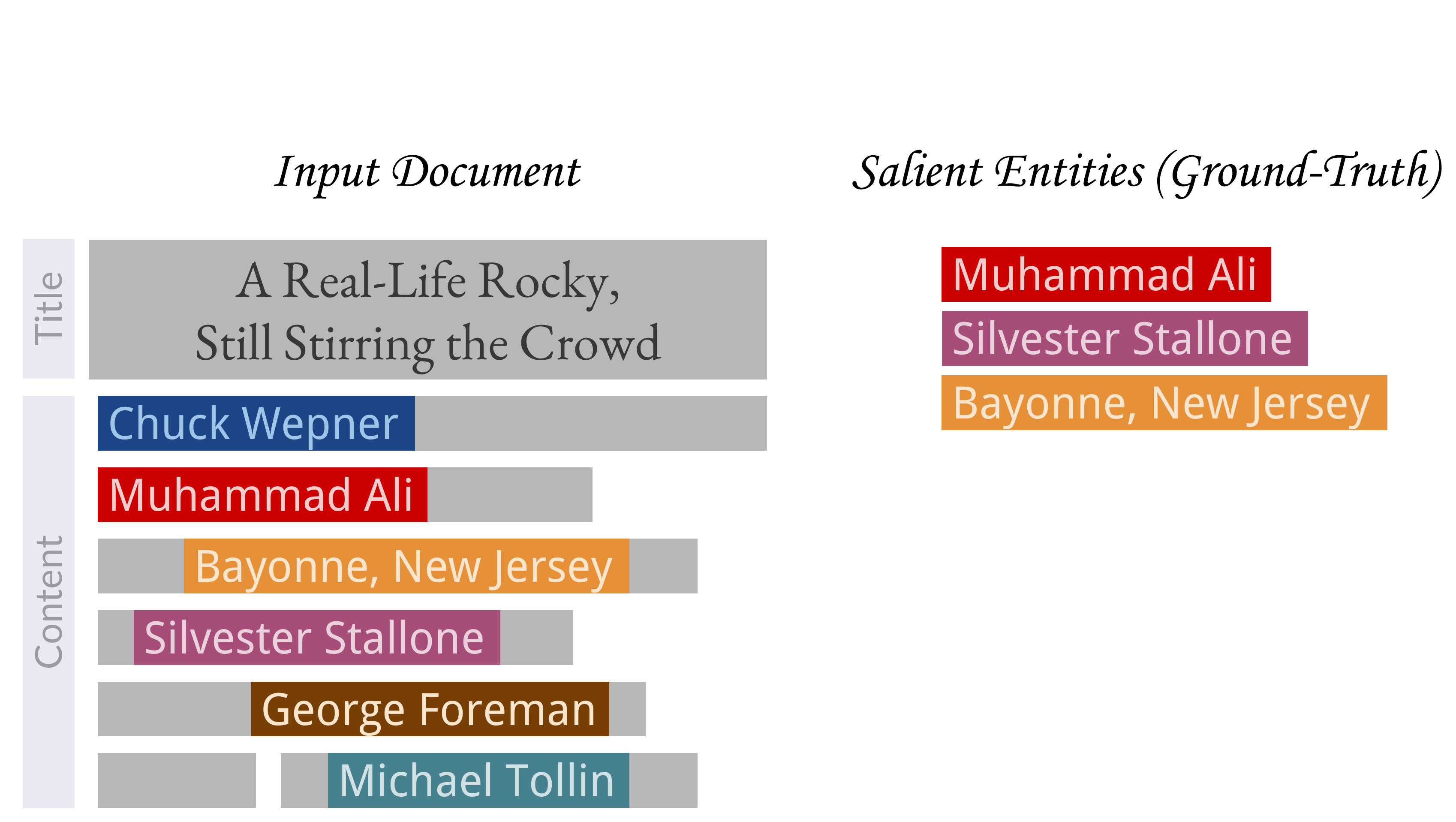}
				\vspace{0.2cm}
				\caption{
					Example where the latent information between the title of the input document and the candidate entities can help in distinguishing between salient and non-salient entities. In fact, despite \textsf{Silvester Stallone} and \textsf{Muhammed Ali}  do not explicitly appear in the document's title, their relationship with Rocky is a clear strong indication of their salience. }
				\label{fig:features-title}
			\end{figure}

		\bigskip\noindent\textbf{Relatedness-based Features.}
		These features are introduced to capture how much an entity $e$ is related to all other entities in the input document $d$, with the intuition that if an entity is salient then its topic should not be \textit{isolated} in $d$. \textsc{Swat} uses two main groups of relatedness functions~\citep{Ponza:2017:TFC:3132847.3132890}: 
		\begin{enumerate}
			\item the Jaccard relatedness described by~\cite{piccinno2014tagme}, since its deployment in the disambiguation phase of \textsc{Wat} achieves the highest performance over different datasets~\citep{usbeck2015gerbil};
			
			\item the cosine similarity between the latent embeddings of the compared entities, either based on Entity2Vec or on DeepWalk models.
		\end{enumerate}
		
		\noindent Furthermore, we use these two relatedness functions  in order to compute two more classes of features: 
		\begin{enumerate}
			\item the ones based on several centrality algorithms --- i.e., Degree, PageRank, Betweenness, Katz, HITS, Closeness and Harmonic~\citep{boldi2014axioms} --- applied over three versions of the entity graph described in Stage 1. We recall that this is a complete graph where nodes are entities and  edges are weighted with a similarity measure between the connected entities which is estimated either with Jaccard, or with Entity2Vec, or with DeepWalk.
			
			\item  the ones based on proper statistics aggregating the relatedness scores between the entity $e$ and other entities in $d$.
		\end{enumerate}

			\smallskip\textit{Captured Phenomena.} Through these features we aim at capturing how much an entity is semantically central with respect to the other annotated entities. Figure~\ref{fig:features-pagerank} shows an intuitive example where the centrality of entities clearly play a role in discriminating between salient and non-salient entities. More precisely, highly related entities receive higher centrality scores (i.e., \textsf{New York City} and \textsf{Fashion Week}), whereas the ones that are poorly related with the others (i.e., \textsf{Lower East Side}) receive lower centrality scores and hence should be classified as less salient for the content of the input document.

			\begin{figure}[t]
				\centering
				\includegraphics[scale=0.25]{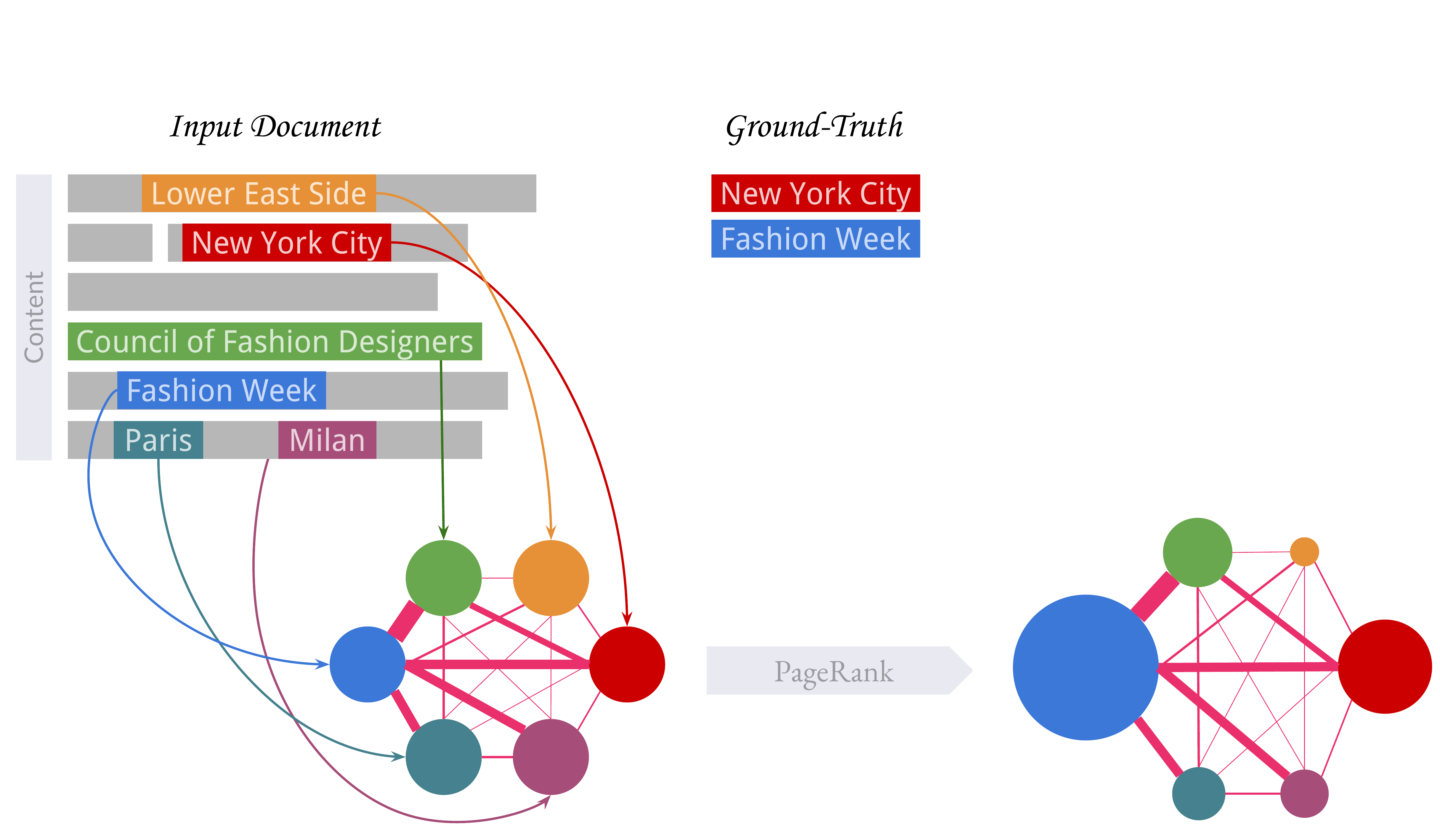}
				
				\vspace{0.2cm}
				\caption{
					A graph where nodes are entities annotated in the input document and edges are weighted with the relatedness score between the connected entities. Ticker edges indicate higher weights. Centrality scores are eventually computed by running PageRank, and they should show how the relatedness-based feature can help in distinguishing salient and non-salient entities.}
				\label{fig:features-pagerank}
			\end{figure}

			\noindent According to \cite{boldi2014axioms}, centrality can actually be defined in several ways and literature currently does not offer a uniform terminology as well as different centrality algorithms capture different aspects of nodes and their connections in the graph. Degree offers a ``majority voting'' between nodes, PageRank computes the probability that a random surfer passes into a node by intermittently teleporting back to other nodes,  Betweenness measures the volume of the shortest paths passing through a given node,  Katz sums the weighted paths coming into a node, HITS scores each node with a high authoritative (resp. hub) value whether the node at hand is pointed by many good hub (resp. authoritative node), Closeness assigns a higher score to nodes that have smaller distance with respect to all the others in the graph, and Harmonic measures the harmonic mean of all distances between every pair of nodes. Because it is unknown what kind of centrality algorithms could be more effective in the context of entity salience, we decided to investigate the use of all of them over the graph of entities described above.

			\vspace*{0.5cm}

		\section{Validation Methodology}
		\label{sec:exp}
		
		In this section, we describe the validation methodology adopted for the evaluation our system \swat{}. Section~\ref{sub:datasets} describes the datasets used in our benchmarks by reporting the main differences between the two test-beds, Section~\ref{sub:classification-methods} describes the experimented tools whose results are discussed in Section~\ref{sec:discussion}.

		\subsection{Datasets} \label{sub:datasets}
		
		The experimental validation of the accuracy and efficiency performance of \swat{} is executed on the following datasets.

			\smallskip\noindent\textbf{New York Times.} The annotated version of this dataset, suitable for the entity salience problem, was introduced by~\cite{dunietz2014new}. It consists of annotated news drawn from 20 years of the New York Times newspaper~\citep{sandhaus2008new}. It is worth to point out that the numbers reported by~\cite{dunietz2014new} are slightly different from the ones we derived by downloading this dataset: authors informed us that this is due to the way they have exported annotations in the final release and this impacts onto the F1-performance of their system for about $-0.5\%$ in absolute micro-F1. We will take these figures into account in the next sections when comparing \textsc{Swat} with the \textsc{Cmu-Google} system.
			
			Since the entity linker used by~\cite{dunietz2014new} is not publicly available (and this was used to derive the ground truth of the NYT dataset), we kept only those entities which have been annotated by both \textsc{Swat} and \textsc{Cmu-Google}. The final figures are the following: the news in the training+validation set are $99,348 = 79,462 + 19,886$, and are $9,577$ in the test set; these news contain a total of $1,276,742$ entities in the training+validation set (i.e., $1,021,952 + 254,790$) and $19,714$ entities in the test set. Overall the  dataset contains $108,925$ news, with an average number of 975 tokens per news, more than 3 million mentions and $1,396,456$ entities, of which $14.7\%$ are labeled as salient.

			\smallskip\noindent\textbf{Wikinews.} This dataset was introduced by~\cite{trani2016sel}, it consists of a sample of news published by Wikinews from November 2004 to June 2014 and annotated with Wikipedia entities by the Wikinews community. This dataset is significantly smaller than NYT in all means: the number of documents ({365} news), their lengths (an average of {297} tokens per document) and number of annotations (a total of $4,747$ manual annotated entities, of which $10\%$ are labeled as salient). Nevertheless, this dataset has some remarkable features with respect to NYT: the ground-truth generation of the salient entities was obtained via human-assigned scores rather than being derived in a rule-based way, and it includes both proper nouns (as in NYT) and common nouns (unlike NYT) as salient entities. For the cleaning of the dataset, we follow \citep{trani2016sel} as done in their experimental setup by removing the 61 documents that do not have any salient entity.

		As far as the dataset subdivision and evaluation process are concerned, we use the following methodology. For the NYT, we use the same training/testing splitting as defined by~\cite{dunietz2014new} as detailed above, whereas for Wikinews we deploy the evaluation procedure described by~\cite{trani2016sel}, namely the averaged macro-F1 of a 5-fold cross-validation.

		\subsection{Tools}
		\label{sub:classification-methods}
		
		\noindent\textbf{Baselines.} We implement four baselines. The first one is the same baseline introduced by~\cite{dunietz2014new}, which simply classifies an entity as salient if it appears in the first sentence of the input document. The other three baselines are new and try to investigate the individual power of several novel features adopted by \textsc{Swat}. More precisely, the second baseline (i.e., $\rho$-baseline) extends the previous one by adding the check whether the $\rho$-score (capturing entity coherence) is greater than a fixed threshold. The third (resp. fourth) baseline classifies an entity as salient if its maximum TextRank (resp. Rel-PageRank) score is greater than a fixed threshold.

		\smallskip
		\noindent \textbf{Two Versions of the \textsc{Cmu-Google} System.} The original system~\citep{dunietz2014new} uses a proprietary entity linker to link proper nouns to Freebase entities, and then classify them into salient and non-salient by deploying a small number of standard features based on position and frequency. This system is not available to the public, thus we  report in our tables the performance figures published by \cite{dunietz2014new}.
		
		Furthermore, in order to support experiments over the new dataset Wikinews, we decide to implement our own version of the \textsc{Cmu-Google}'s system by substituting the proprietary modules with open-source tools: we used \textsc{Wat} as entity linking system \citep{piccinno2014tagme} and a state-of-the-art logistic regressor as classifier \citep{pedregosa2011scikit}. Our (re-)implementation achieves performance very close to the original system (see Table \ref{table:nytresults}) and thus it is useful to obtain a fair comparison over the Wikinews dataset.

		\smallskip \noindent \textbf{The \textsc{Sel} System.} This is the system proposed by~\cite{trani2016sel} that uses a machine learning regressor to detect salient entities via a set of features that is wider than the ones used in \textsc{Cmu-Google}. This system is not available to the public, so we report in our tables the performance figures published by \cite{trani2016sel}.

		\smallskip \noindent \textbf{Configurations of \textsc{Swat} and Baselines.}  We experiment different configuration settings of \textsc{Swat} and of the baselines above, according to the characteristics of the ground-truth datasets.  For NYT, where the ground-truth was generated by assuming that salient entities can be mentioned in the text only as proper nouns, we configured these systems to annotate only proper nouns detected by {CoreNLP}; whereas for Wikinews, where the ground truth comes with no assumptions, we tested two variants: one detecting only proper nouns, and the other detecting both proper and common nouns. For the tuning of {XGBoost}'s classifier, we performed a grid-search over typical values of its hyper-parameters,  finding the \textit{best values} (i.e., the ones performing better on the validation sets of New York Times and Wikinews, respectively) reported in Table \ref{table:hyperparameters}.

				\begin{table}[t!]
					\caption{Candidate values  and the best configuration found by the grid-search procedure for the tuning of {XGBoost}'s hyper-parameters on New York Times and Wikinews datasets.}
					\label{table:hyperparameters}
					
					\resizebox{\textwidth}{!}{
						\begin{tabular*}{1.1\textwidth}{@{\extracolsep\fill}llrr}
							\toprule
							\textbf{Hyper-parameters} & \textbf{Candidate Values} & \textbf{New York Times}   & \textbf{Wikinews} \\ \midrule
							\textsf{max\_depth}             & \{2, 4, 6,  8\}                          & 8                & 2         \\
							\textsf{min\_child\_weight}     & \{6, 8, 10\}                            & 6                & 6         \\
							\textsf{gamma}                  & \{0.1, 0.3, 0.5\}                       & 0.1              & 0.5       \\
							\textsf{reg\_alpha}             & \{0.001, 0.01, 0.05\}                  & 0.001            & 0.05      \\
							\textsf{scale\_pos\_weight}     & \{1, 2, 3, 4, 5, 6, 7, 8, 9, 10\}               & 2                & 8         \\ \hline
						\end{tabular*}
					}
				\end{table}

		\section{Analysis and Discussion}
		\label{sec:discussion}
		
		We first experiment our proposed solution \swat{} against the state-of-the-art over the two datasets New York Times and Wikinews (Section~\ref{sub:exp-results}). Then, we analyze and discuss several aspects of our proposed system by focusing on: (i) the generalization ability of the tested systems as a function of the used training data (Section~\ref{sub:gen-ability}), (ii) the dependence between the size of the training set and the accuracy of our solution (Section~\ref{sub:acc-vs-trainsize}), (iii) the impact that features have on the quality of the predictions (Section~\ref{sec:feature-analysis}), (iv) the time efficiency of \swat{} according to its main components and its overall speed-up when only the most relevant features are used (Section~\ref{sec:eff}), (v) the dependence of top-systems on the position of the salient entities within the input document, and (vi) an analysis of the limitations of the current systems in terms of the types of erroneous predictions (Section~\ref{sub:error-analysis}).

		\subsection{Experimental Results}
		\label{sub:exp-results}
		
		Experimental figures on the two datasets are reported in Tables~\ref{table:nytresults}--\ref{table:wikinewsresults}, where we denote by \textsc{Cmu-Google-ours} our implementation of the system by~\cite{dunietz2014new}. This system is only slightly worse than the original one, which could depend on the differences in the NYT dataset commented above and in the deployment of open-source modules rather the Google's proprietary ones. The final performance of \textsc{Cmu-Google-ours} is very close to what claimed by~\cite{dunietz2014new}, thus we decide to use this software also on the Wikinews dataset.  We notice that both TextRank and Rel\mbox{-}PageRank baselines obtain low micro- and macro-F1 performance over both datasets. This  is probably due to the characteristics of these datasets: the salient information in news is typically confined to initial positions, so those systems are drastically penalized by ignoring positional information. This statement is further supported by the results of Positional and Positional\mbox{-}$\rho$ baselines: they are trivial but generally achieve better performance.

		\begin{table*}[t]
			
			\centering
			\caption{Performance of the tested systems on the New York Times' dataset. Statistically significant improvements are marked with $^\blacktriangle$ for $p < 0.01$. \label{table:nytresults}}
			
			\resizebox{\textwidth}{!}{
				\begin{tabular*}{\textwidth}{@{\extracolsep\fill}@{}lcccccc@{\extracolsep\fill}}
					\toprule
					\multirow{2}{*}{\textbf{System}} & \multicolumn{3}{@{}c@{}@{}}{\textbf{Micro}} & \multicolumn{3}{@{}c@{}@{}}{\textbf{Macro}}  \\ \cmidrule{2-4}  \cmidrule{5-7}

					& \textbf{Precision}               & \textbf{Recall}               & \textbf{F1}              & \textbf{Precision}               & \textbf{Recall}               & \textbf{F1}   \\ \hline
					\addlinespace Positional Baseline                                & 59.1            & 38.6            & 46.7            & 39.0            & 32.7            & 33.0\\
					Positional-$\rho$ Baseline                         & 61.9            & 36.9            & 46.2            & 38.5            & 31.0            & 32.0 \\
					TextRank                                           & 27.0            & 58.8            & 37.0            & 30.0            & 48.6            & 33.4                      \\
					Rel-PageRank                                       & 20.3            & 62.5            & 30.6            & 21.3            &\textbf{55.3}           & 28.0                      \\
					\textsc{Cmu-Google} & 60.5            & 63.5 & 62.0            & --            & --            & --                      \\
					\textsc{Cmu-Google-ours}                         & 58.8            & 62.6            & 60.7            & 47.6            & 50.5            & 46.1                      \\
					\textsc{Swat}                                   & \textbf{62.4}$^\blacktriangle$ & \textbf{66.0}$^\blacktriangle$            & \textbf{64.1}$^\blacktriangle$ & \textbf{50.7}$^\blacktriangle$ & 53.6 & \textbf{49.4}$^\blacktriangle$           \\ \bottomrule
					
				\end{tabular*}
			}
		\end{table*}

		Table \ref{table:nytresults} reports the results for the experiments on the New York Times dataset. We notice that the new features adopted by \textsc{Swat} allow it to outperform \textsc{Cmu-Google-ours} by $3.4\%$ and $3.3$\% over micro- and macro-F1, respectively, and \textsc{Cmu-Google} by $2.6\%$ in micro-F1 (macro-F1 was not evaluated by \cite{dunietz2014new}). We tested statistical significance with respect to  \textsc{Cmu-Google-ours}\footnote{Since the original  \textsc{Cmu-Google} system is not available we cannot test statistical significance with~respect~to~it.} using a two-tailed paired t-test and we found that all the improvements reported by \swat{} in Table~\ref{table:nytresults} are statistically significant with $p < 0.01$.

		\begin{table*}[t!]
			\caption{Performance on the Wikinews dataset. For each system we report the score obtained by the system configured to annotate either only proper nouns (top) or both proper and common nouns (down).\label{table:wikinewsresults}}
			\centering
			\resizebox{\textwidth}{!}{
				\begin{tabular*}{\textwidth}{@{\extracolsep\fill}@{}lcccccc@{\extracolsep\fill}}
					\toprule
					
					\multirow{2}{*}{\textbf{System}} & \multicolumn{3}{@{}c@{}@{}}{\textbf{Micro}} & \multicolumn{3}{@{}c@{}@{}}{\textbf{Macro}}  \\ \cmidrule{2-4}  \cmidrule{5-7} 
					& \textbf{Precision}               & \textbf{Recall}               & \textbf{F1}              & \textbf{Precision}               & \textbf{Recall}               & \textbf{F1}              \\ \hline
					
					\addlinespace \multirow{2}{*}{Positional Baseline} & 23.3	& 	67.0& 		35.0	& 	25.2	& 	67.0	& 	34.0 \\
					& 14.4	& 	\textbf{72.0}	& 	24.0	& 	16.1	& 	 \textbf{72.7}	& 	25.0\\ \hline

					\addlinespace \multirow{2}{*}{Positional-$\rho$ Baseline} & 36.8	&	60.3	&	45.7	&	38.3	&	61.6	&	43.5\\
					&34.1&		58.5	&	43.1	&	36.2	&	61.3	&	41.9\\ \hline

					\addlinespace \multirow{2}{*}{TextRank}      	&          12.2		&	47.5		&	19.4		&	14.1		&	49.3		&	20.2 \\
					&5.7		&	49.2		&	10.1		&	6.3		&	50.9		&	10.6        \\ \hline
					
					\addlinespace \multirow{2}{*}{Rel-PageRank}        	&          10.0	  	&  	51.0	  	&  	16.8	  	&  	10.1	  	&  	51.2	  	&  	15.9 \\
					&  	10.6	  	&  	35.8	  	&  	16.4	  	&  	11.1	  	&  	34.8	  	&  	14.7   \\ \hline

					\addlinespace \multirow{2}{*}{\textsc{Cmu-Google-ours}} 	& 41.0		& 	60.0		& 	49.0		& 	42.3		& 	61.0	& 		46.0 \\
					& 41.0		& 	56.0		& 	47.0	& 		41.0		& 	58.0		& 	45.0\\ \hline

					\addlinespace\textsc{Sel}   & --          & --       & --            &  \textbf{61.0}           &  50.0           & 52.0            \\ \hline
					
					\addlinespace \multirow{2}{*}{\textsc{Swat}}  &  \textbf{58.0}	 &	64.9	 & \textbf{61.2}	 &	57.7 &		 67.0 &	 \textbf{58.3} \\
					&51.0	 &	67.4	 &	58.0	 &	53.7	 &	69.7	 &	56.6 \\\bottomrule
				\end{tabular*}
			}

			\vspace*{.8cm}
			\caption{Generalization ability of \textsc{Swat} trained on NYT and tested on Wikinews. For each system we report the score obtained by the system configured to annotate either only proper nouns (top) or both proper and common nouns (down).
				\label{table:flexresults}}
			\centering
			\resizebox{\textwidth}{!}{
				\begin{tabular*}{\textwidth}{@{\extracolsep\fill}@{}lcccccc@{\extracolsep\fill}}
					\toprule
					
					\addlinespace \multirow{2}{*}{\textbf{System}} & \multicolumn{3}{@{}c@{}@{}}{\textbf{Micro}} & \multicolumn{3}{@{}c@{}@{}}{\textbf{Macro}}  \\ \cmidrule{2-4}  \cmidrule{5-7} 
					& \textbf{Precision} & \textbf{Recall} & \textbf{F1} & \textbf{Precision} & \textbf{Recall} & \textbf{F1} \\  \hline

					\addlinespace\multirow{2}{*}{\textsc{Swat-clf}} & 35.0	  & 	72.0& 		47.1	& 	37.9	& 	73.7& 		46.7 \\
					& 27.3	& 	\textbf{75.7}	& 	40.1	& 	31.3	& 	\textbf{78.0}	& 	41.5\\ \hline
					
					\addlinespace\multirow{2}{*}{\textsc{Swat-reg}} & \textbf{55.9}	& 	59.9	& 	\textbf{57.7}	& 	\textbf{54.0}	& 	62.4	& \textbf{54.3} \\
					&  49.3	& 	63.1 & 		55.1	& 	50.6	& 	65.9	& 53.3	\\ \bottomrule

				\end{tabular*}	
			}
							\vspace{0.4cm}
			
		\end{table*}

		Table \ref{table:wikinewsresults} reports the results on Wikinews dataset. It goes out without saying that the improvement achieved by \textsc{Swat}  against the state-of-the-art is even larger than on NYT. Specifically, \textsc{Swat} improves the micro-F1 of $12.2\%$ with respect to \textsc{Cmu-Google-ours} and the macro-F1 of  $6.3\%$ with respect to \textsc{Sel}.

		\subsection{Generalization Ability of \textsc{Swat} Trained on NYT}  
		\label{sub:gen-ability}
		
		The second question we experimentally investigate is about the \textit{generalization ability} of the feature set used by \textsc{Swat} varying the dataset on which the training and tuning phases are performed. In particular, we experiment on two different configurations of our system. \textsc{Swat-clf}  is \swat{} trained over NYT and directly used over Wikinews; and \textsc{Swat-reg} is \swat{} trained over NYT but whose regressor is tuned over Wikinews by maximizing the macro-F1 over the training folds.

		According to Table \ref{table:flexresults}, \textsc{Swat-csf} obtains performance lower than the systems specifically trained over Wikinews, such as  \textsc{Swat}  and \textsc{Sel}, but it turns actually to be slightly better than \textsc{Cmu-Google-ours} by +$0.7$\% in macro-F1.

		On the other hand, the tuning on Wikinews by \textsc{Swat-reg} allows our system to achieve better performance in macro-F1 than both \textsc{Cmu-Google-ours} and \textsc{Sel}: +$8.7\%$  in micro-F1 with respect to \textsc{Cmu-Google-ours} and of +$8.3$\% and +$2.3$\% in macro-F1 with respect to \textsc{Cmu-Google-ours} and \textsc{Sel}.
		These figures show that the features introduced by \textsc{Swat} are flexible enough to work independently from the news source and without overfitting the single-source training data (i.e., NYT).

		\vspace{0.4cm}
		\subsection{Accuracy versus Training Size}
		\label{sub:acc-vs-trainsize}

		We analyze the performance of the two versions of \textsc{Swat} with respect to different sizes of the training data. We focus these experiments on the largest dataset available, namely New York Times. 
		
		Figure~\ref{fig:trainsize} provides a side-by-side comparison of the performance of the two systems when $5\%$, $25\%$, $50\%$, $75\%$ and $100\%$ of the whole training data is used. The original validation set is kept for the tuning of the hyper-parameters, as described in Section \ref{sec:exp}. Micro-precision, -recall and -F1 are finally calculated over the test-set.
		
		The precision of \swat{} increases until when $50\%$ of the whole training size is used, with a peak of $63.5\%$. Unfortunately, when more than $50\%$ of the training data is used, the precision decreases by eventually losing $-1.1\%$ in performance. This degradation is due to the increase of the recall that eventually allows \swat{} to consistently improve its micro-F1 until the whole training set is used.
		
		\begin{figure}[t!]	
			\centering
			\includegraphics[width=\textwidth]{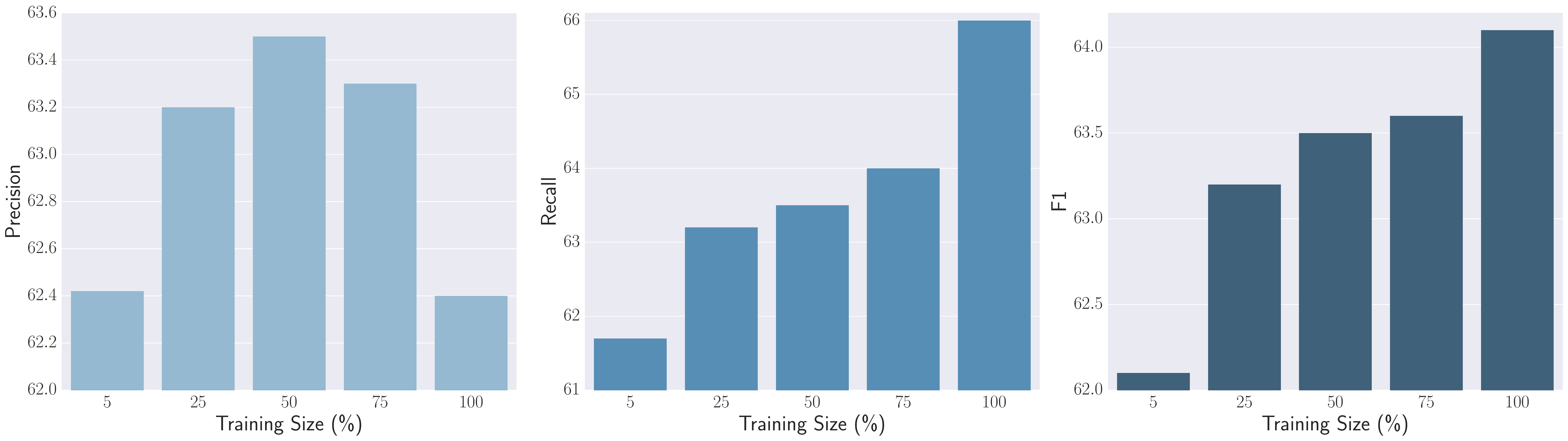}
			\caption{Comparison of the performance \textsc{Swat} over different training sizes of the New York Times dataset.
				\label{fig:trainsize}}
			
		\end{figure}

		\newpage
		\subsection{Feature Analysis}\label{sec:feature-analysis}
		
		\begin{figure}[t!]
			\centering
			
			\includegraphics[scale=0.32]{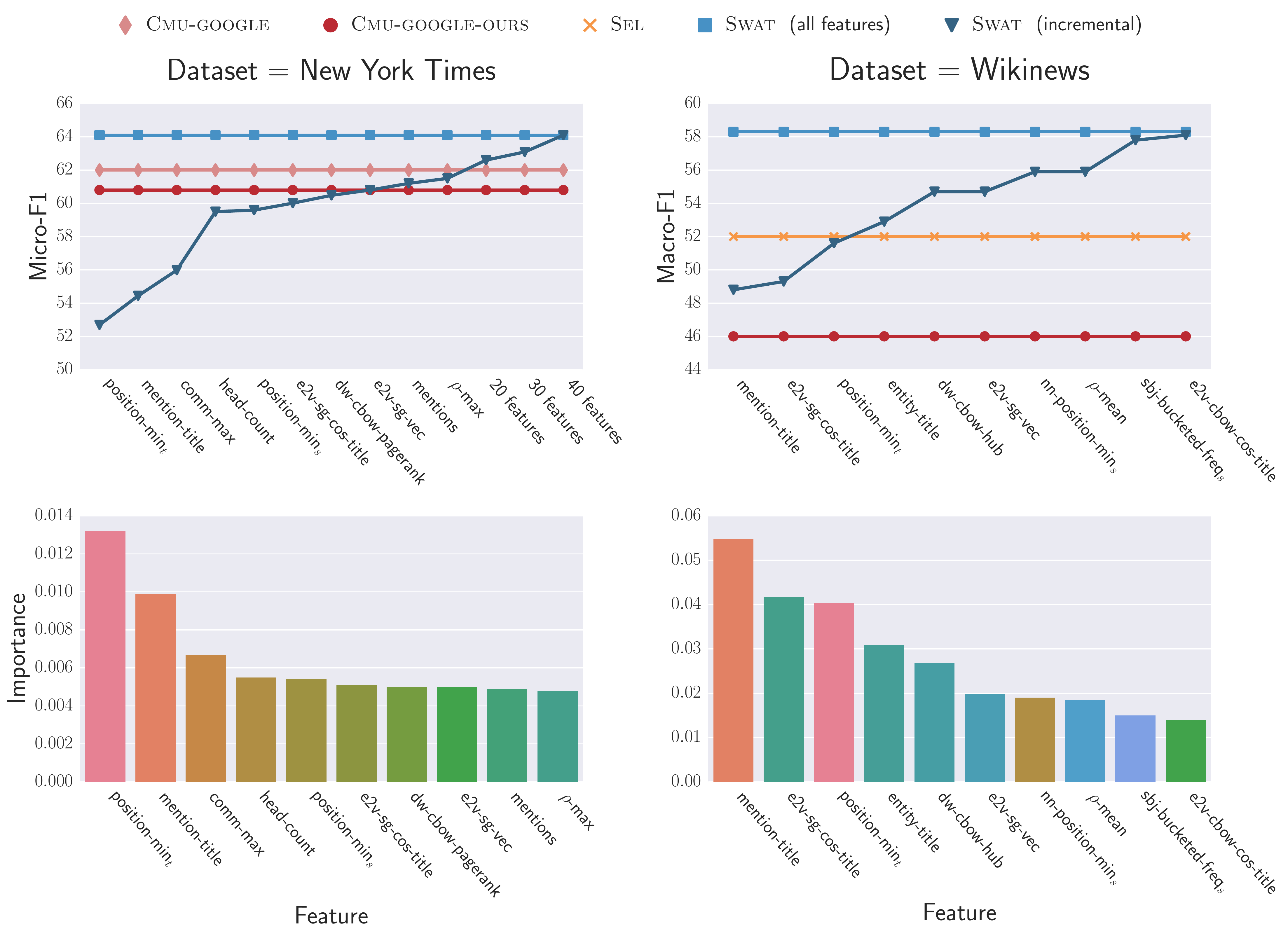}
			
			\caption{\label{fig:swat2.0incfs} Performance of the incremental feature addition (top) of \textsc{Swat} according to the corresponding feature importance provided by {XGBoost} (bottom) over NYT (left) and Wikinews (right) datasets.} 
		\end{figure}

		Let us jointly discuss the most important signals emerging from the incremental feature additions experimented with  \textsc{Swat} on both datasets (see Figure~\ref{fig:swat2.0incfs}). Through this analysis, we aim to clarify what are the key elements needed for the entity salience detection. 
		
		We notice that the most important features for our system depend on four common elements: (i) position (e.g., $position\mbox{-}min_t$), (ii) the latent similarity between an entity and the title (e.g., \textit{e2v-sg-cos-title}), (iii) the centrality of an entity (e.g., $dw\mbox{-}$ $cbow\mbox{-}pagerank$ and $dw\mbox{-}cbow\mbox{-}hub$) and finally the (iv) coherence scores of the annotated entities (e.g., $comm\mbox{-}max$ and $\rho\mbox{-}mean$). On the other hand, frequency signals are fundamental when the input document is large, such as in the NYT dataset (e.g., $head\mbox{-}count$ or $mentions$), whereas on relatively shorter texts, such as in  Wikinews, they are less useful and they bring improvements only when combined with other signals, such as dependency and positional information (e.g., $sbj\mbox{-}$$bucketed\mbox{-}$$freq_s$).

			We mention here that during this analysis we found several novel errors that are committed by \swat{} despite its results being better than \textsc{Cmu-Google} system. More precisely, it is very common that an entity that is salient it is present at the beginning of the document, whereas if it appears too far it is a common practice to classify it as non-salient. Nevertheless, we found some cases where the features we designed overcome these problems. Accordingly, we report here several examples where it is evident at human inspection that the designed feature helps the system in improving its predictions thus showing where our system's predictions mainly differ from \textsc{Cmu-Google}. 
			
			For the ease of explanation, we report the whole ground-truth and predictions for both systems, while since input documents are very large we report only several but meaningful annotated entities.

			\bigskip\noindent\textbf{Qualitative Comparison between \swat{} and \textsc{Cmu-Google}.} In this paragraph we aim at shading more light into \textit{how} the new feature space that we designed for \swat{} allows our system to achieve higher-quality predictions than \textsc{Cmu-Google}'s ones. In accordance with the best features identified by {XGBoost}, we report here several practical examples of frequent patterns that we have identified during our analysis and that explicitly show where our \textit{new} and \textit{most relevant} features help \swat{} in achieving better performance than  \textsc{Cmu-Google} system. In this analysis, we did not consider frequency-based  features (i.e., $head\mbox{-}count$ and $mentions$) since they are equivalent to the ones already proposed and used by the \textsc{Cmu-Google} system.  For ease the understanding of these common patterns, we structured our graphical examples (in Figures \ref{fig:example:position}, \ref{fig:example:title}, \ref{fig:example:rho}, \ref{fig:example:pagerank}) as follows. On the left, we report a meaningful  subset of entities annotated in the input text (since position is a very strong feature, we preserve the order of the annotated entities), in the center we distinguish the two systems (\swat{} and \textsc{Cmu-Google}, respectively) and, finally, on the right we report the whole set of predicted salient entities as well as the ground-truth.

			\medskip\noindent\textit{Position-based Features.} As expected, the new features designed with token-level granularity (i.e., $position\mbox{-}min_t$) allow \swat{} to achieve a better quality in the detection of salient entities. More precisely, when an entity is mentioned at the beginning of the document (but not in the first few sentences) it is commonly classified by \textsc{Cmu-Google} as non-salient since it obtains a large value for $1st\mbox{-}loc$. Figure~\ref{fig:example:position} shows an example where a salient entity is mentioned at the beginning but, since it appears for the first time only in the third sentence, it is classified by the \textsc{Cmu-Google} system as non-salient. This situation repeats frequently in the experimental datasets.
			
			\begin{figure}[t]
				\centering
				\includegraphics[width=1.\textwidth]{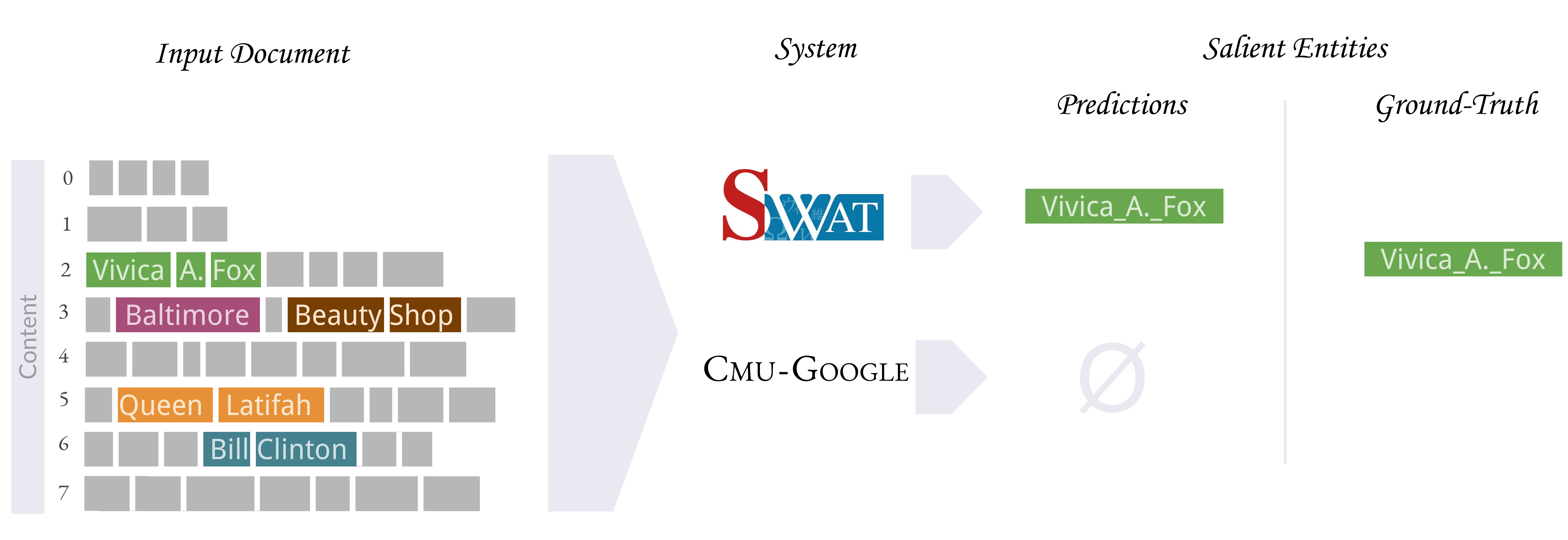}
				\caption{Example that shows where token-level features allows \swat{} to detect the position of a salient entities as at the beginning of a document despite it does not appear in the very first sentences.  }
				\label{fig:example:position}
			\end{figure}
			
			On the other hand, the use of a token-level feature allows \swat{} to annotate \textsf{Vivica A. Fox} at the very beginning and properly classify it as salient. From our analysis, we found that working at token-level makes our system more robust than \textsc{Cmu-Google}, which actually works at sentence-level. Token-level features are more flexible, especially in the cases where the document has several small sentences at the beginning, which induce $1st\mbox{-}loc$ easily to get large values, as opposite to $position\mbox{-}min_t$, which keeps its score low also in these cases.

			\medskip\noindent\textit{Title-based Features.} In our system we introduced two different title-based features (i.e., $mention\mbox{-}title$ and $e2v\mbox{-}sg\mbox{-}cos\mbox{-}title$) which aim at improving the quality of the entity-salient classification with information coming from the title of a document. The first feature (i.e., $mention\mbox{-}title$) is actually very simple: when an entity is mentioned in the title it is clearly a strong indication of its salience in the document since the author of the news was probably aiming to attract the attention of the reader at first glance. On the other hand, the title can contain information that is related to some entities but without explicitly mentioning them. Nevertheless, \textsc{Swat} is still able to capture these related entities and classify them as salient for the input document. An example of this last case is reported in Figure~\ref{fig:example:title}. 	We notice that both systems predict as salient the entities \textsf{Chuck Wepner} and \textsf{Muhammed Ali} that are mentioned at the beginning of the document; but, in addition, \swat{} is able to correctly detect \textsf{Silvester Stallone} as salient because it is highly related to the title which mentions \textsf{Rocky}, the movie where the actor has played as the main character.

						\begin{figure}[t]
							\includegraphics[width=1.\textwidth]{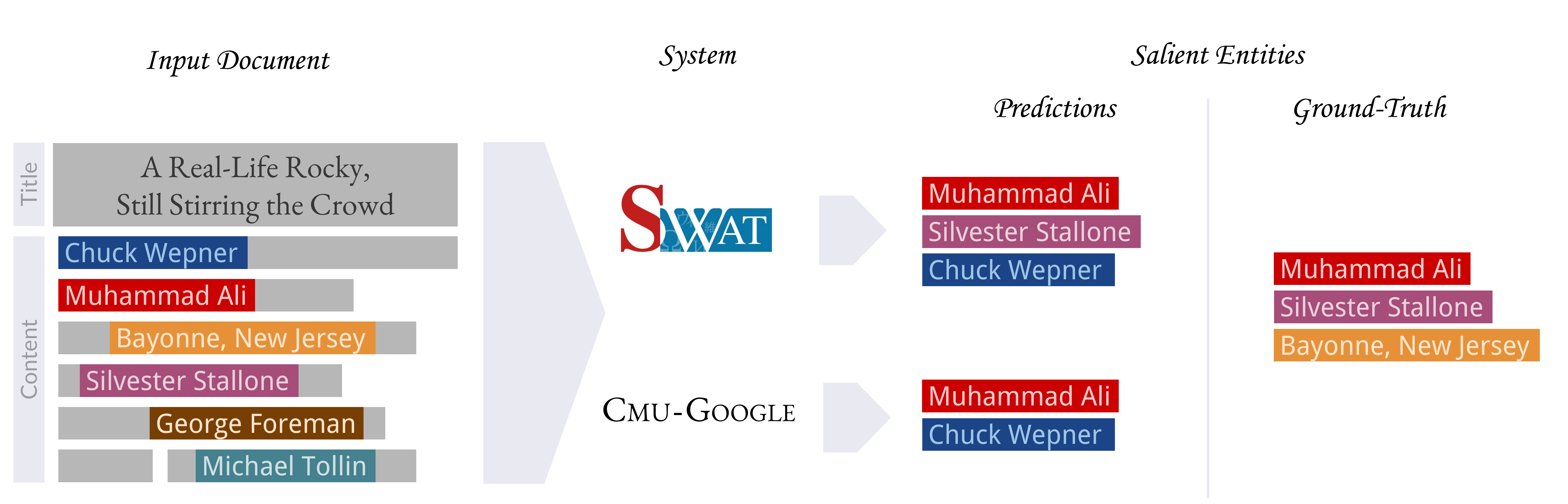}
							\vspace{0.1cm}
							\caption{Example that shows where information present in the title help \swat{} into a proper detection of salient  entities.}
							\label{fig:example:title}
						\end{figure}
						
									\begin{figure}[t]
										\includegraphics[width=1.\textwidth]{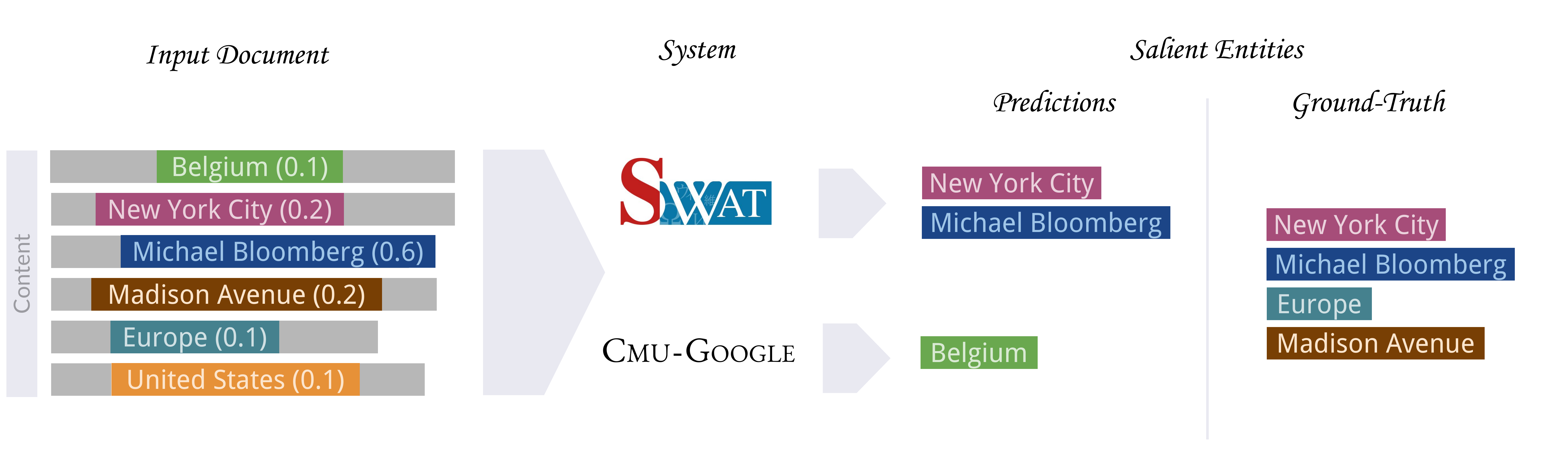}
										\caption{Example that shows the robustness of the \swat{}'s features based on the coherence of the annotations. We show between parenthesis the feature $\rho\mbox{-}max$.}
										\label{fig:example:rho}
									\end{figure}

			\medskip\noindent\textit{Annotation-based Features.} Features based on the scores associated to the annotations (i.e., $comm\mbox{-}max$ and $\rho\mbox{-}max$) make \swat{} even more robust with respect to non-coherent entities. The most interesting case for the proper understanding of the effectiveness of these features is showed in Figure~\ref{fig:example:rho}. For each entity, the $\rho\mbox{-}max$ feature score is reported between parentheses. This example shows a case where an entity can be mentioned at the beginning of the document but without being salient. Unlike {\sc Cmu-Google}, \swat{} is robust in detecting such a kind of entities because the feature $\rho\mbox{-}max$ gets a low score of coherence (in the text, \textsf{Belgium} is an adjective but it is wrongly annotated as the country), which therefore allows to correctly classify it.

						\begin{figure}[t]
							\includegraphics[width=1.\textwidth]{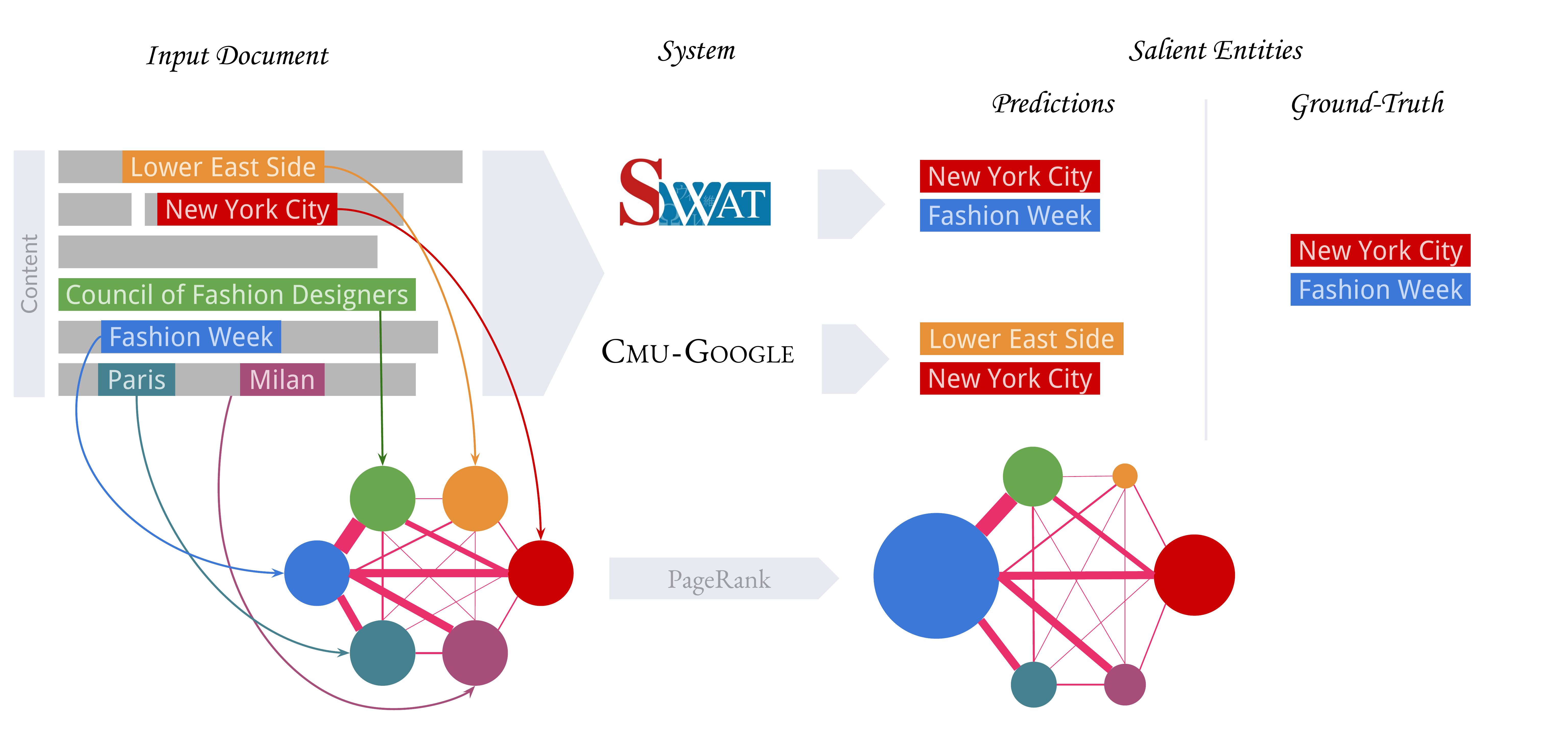}
							\caption{Example that shows where $dw\mbox{-}cbow\mbox{-}pagerank$ feature helps \swat{} in predicting the salient entities. Despite several entities are not mentioned at the very beginning of the input text, they are classified as salient (resp. non-salient) because they achieve high (resp. very low) scores for $dw\mbox{-}cbow\mbox{-}pagerank$. Ticker edges mean higher DeepWalk cosine similarities between two nodes of the graph.}
							\label{fig:example:pagerank}
							\vspace{0.5cm}
						\end{figure}
			
			\medskip\noindent\textit{Relatedness-based Features.} The final set of relevant features that help \swat{} in performing more accurate predictions is represented by the features developed on the top of relatedness signals (i.e., $dw\mbox{-}cbow\mbox{-}pagerank/hub$ features). More precisely, these features contribute to building a complete graph where nodes are the entities annotated in the input document and edges are weighted with the cosine similarity between their DeepWalk embeddings. The relatedness-based features for each entity are eventually computed by running a centrality algorithm (e.g., PageRank) over this graph. These features help \swat{} in predicting as salient those entities which are central with respect to the other entities annotated in the input text. This is especially useful when a salient entity is not mentioned at the beginning of the input text but it is highly related with the rest of the entities present in the input document.

			\smallskip Figure~\ref{fig:example:pagerank} reports a practical example where we show the usefulness of these features, in particular of $dw\mbox{-}cbow\mbox{-}pagerank$. As we can see, both systems predict a correct salient entity that is mentioned at the beginning, namely \textsf{New York City}.   But \textsc{Cmu-Google} classifies as salient also \textsf{Lower East Side} because it appears at the beginning, even if it is not. The reason why our \swat{} does not make this error is that it takes into account how much this entity is very low related to the others.
			
			On the other hand, \swat{} correctly predicts \textsf{Fashion Week} as salient instead of \textsf{Lower East Side}. By carefully looking at the computation of $dw\mbox{-}cbow\mbox{-}pagerank$ feature (bottom of Figure~\ref{fig:example:pagerank}), the node of \textsf{Fashion Week} is linked to the others through heavy weights (ticker edges) than the ones drawn by \textsf{Lower East Side}. More precisely,   \textsf{Fashion Week} has a strong relatedness with \textsf{New York City}, \textsf{Paris} and \textsf{Milan} because they are popular fashion capitals. After the PageRank computation upon this graph, \textsf{Fashion Week} is scored with the highest $dw\mbox{-}cbow\mbox{-}pagerank$ value, whereas \textsf{Lower East Side} is scored much lower. Overall, \swat{} is able to detect as salient entities both \textsf{New York City} (which appears at the beginning of the text) and  \textsf{Fashion Week} (scored with a high $dw\mbox{-}cbow\mbox{-}pagerank$ value).		
			
		}
		
		\subsection{Time Efficiency}\label{sec:eff}
		
		\begin{figure}[t!]
			\centering
			\includegraphics[scale=0.35]{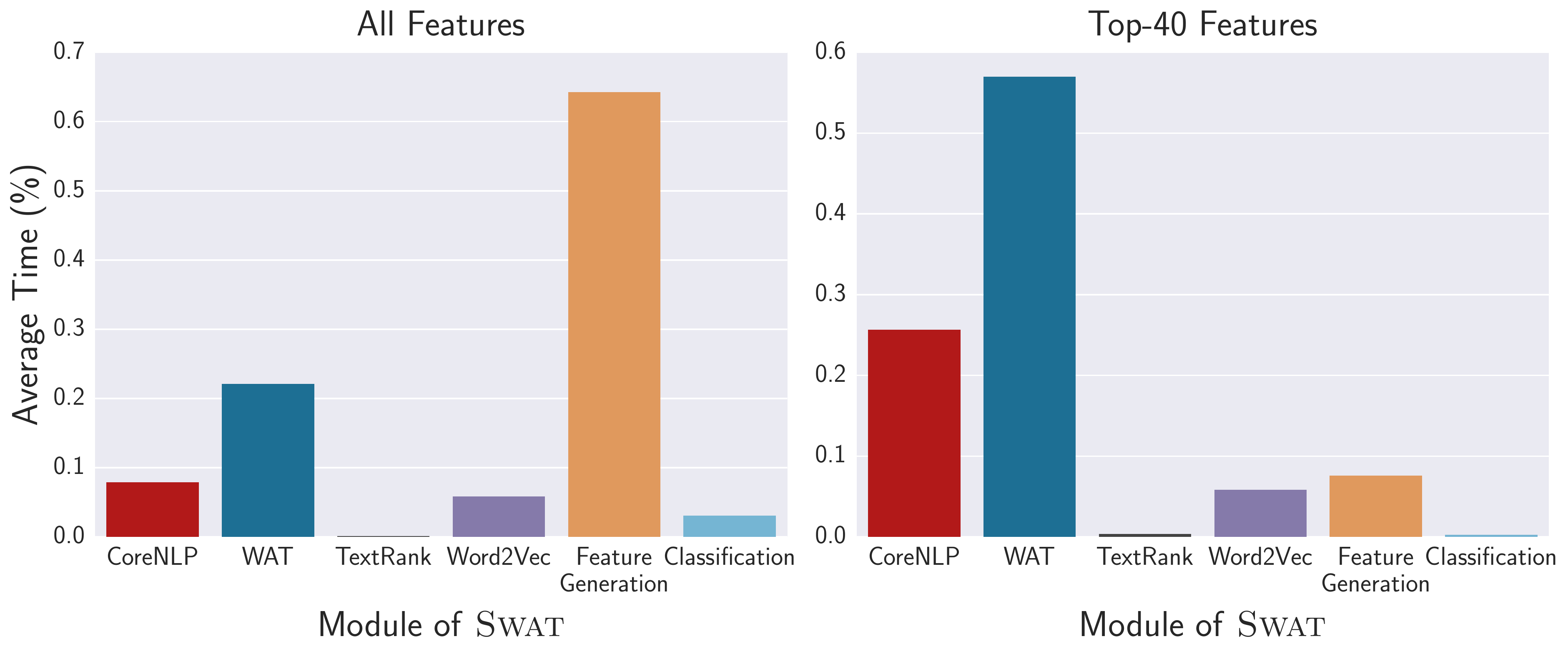}
			\vspace{0.1cm}
			\caption{Average computation time (percentage) of each \textsc{Swat} module for the whole salience-annotation pipeline by deploying all (left) and the top-40 (right) features. Performance is averaged over a sample of 400 documents of the NYT dataset.
				\label{fig:computationtime:perc}}
			
			\vspace{.3cm}
			
			\centering
			\includegraphics[scale=0.225]{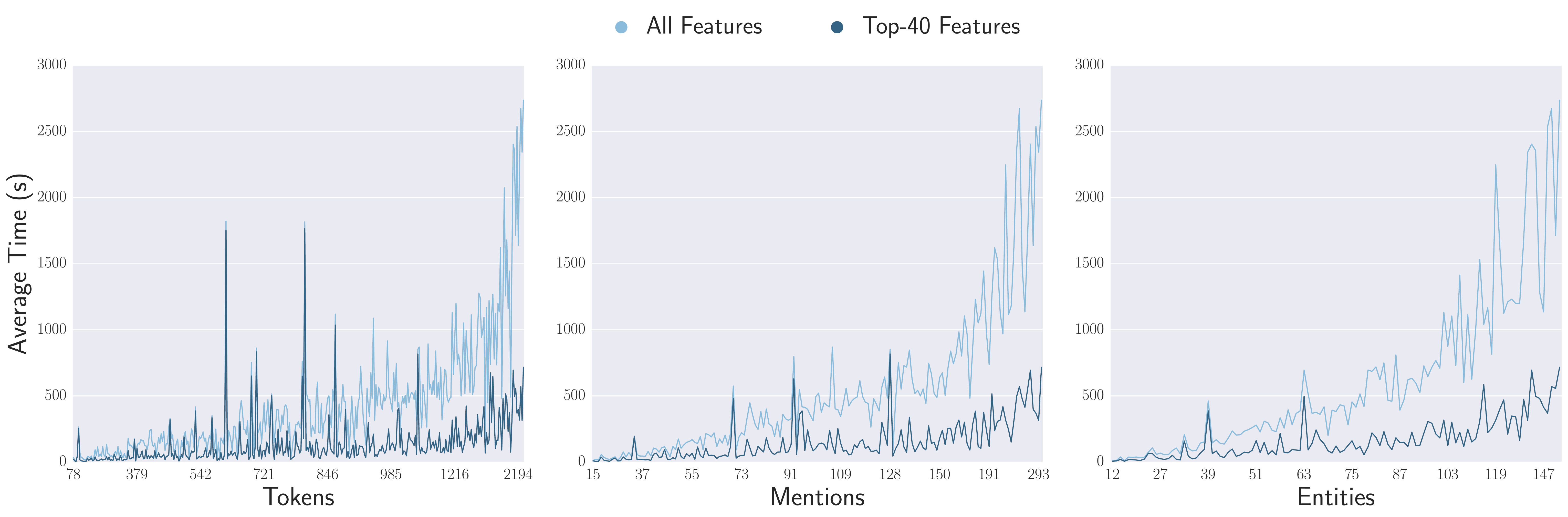}
			\vspace{0.1cm}
			\caption{Average computation time of \textsc{Swat} by distinguishing between the number of tokens, mentions and entities over a sample of 400 documents of the NYT dataset.
				\label{fig:computationtime:time}}

		\end{figure}
		
		The average computation time of each module constituting \textsc{Swat} is reported in Figure~\ref{fig:computationtime:perc}. When all features are used, the most expensive component is clearly the Feature Generation module, which takes about the $64\%$ of the whole computation time of \textsc{Swat}; whereas {CoreNLP}, \textsc{Wat}, {TextRank}, {Word2Vec} and {Classification} take respectively the $7\%$, $22\%$, $0.1\%$, $5.7\%$ and $2\%$ of the computation time of the whole entity-linking-and-salience pipeline. Conversely, when only the top-40 features learned over NYT are used, \textsc{Swat} becomes much faster (up to $5\times$, see Figure~\ref{fig:computationtime:time}) without any significant degradation on its accuracy (see Figure~\ref{fig:swat2.0incfs}). The choice of training \textsc{Swat} over NYT data is motivated by the fact that: (i) the most important features are very similar to the ones derived when the system is trained on Wikinews, and (ii) the system trained on NYT and then tested on Wikinews still obtains higher performance than current state-of-the-art systems (see Section~\ref{sec:exp}).

		\begin{figure}[t]
						
			\centering
			\includegraphics[scale=0.4]{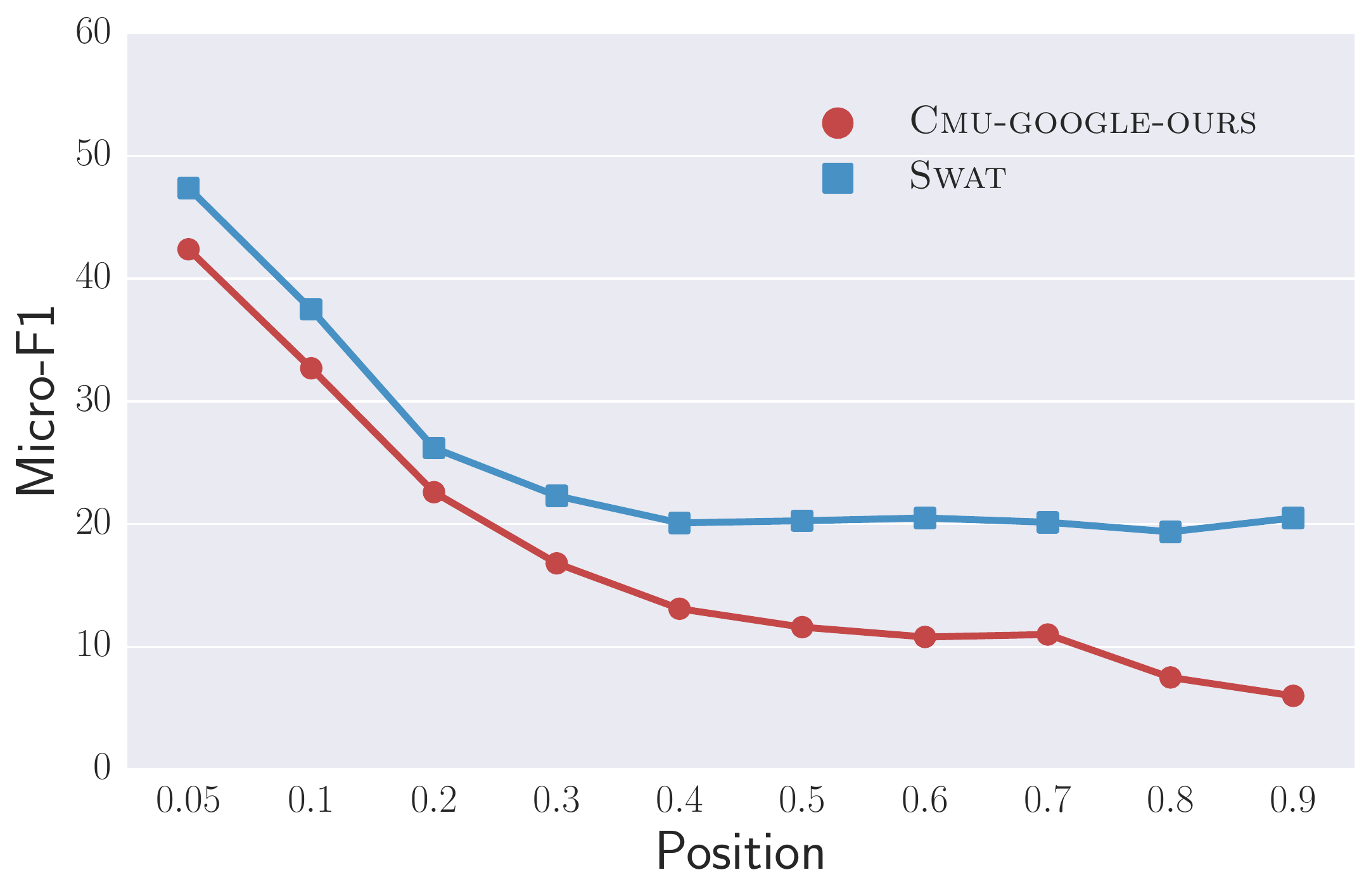}
			\vspace{0.1cm}
			\caption{Micro-F1 performance as a function of the first token positions on the NYT dataset. Each point $(x, y)$ indicates that the micro-F1 is $y$ for all entities whose position is larger than $x$.} \label{fig:f1positions}
		\end{figure}

		\subsection{Flexibility over Entities' Position} 
		\label{sub:flex-entity-pos}
		
		In this section, we address a question posed by~\cite{dunietz2014new} and concerning with the evaluation of how the performance of top-systems depends on the distribution of  the salient entities in the input documents. Figures~\ref{fig:nyt-wn-pos}--\ref{fig:all-nyt-wn-pos} motivated further this question because they show the distribution of the salient and non-salient entities within the NYT and Wikinews datasets. As expected, most of the salient entities are concentrated on the beginning (i.e., position in the first 20\%) of the news over both datasets. Moreover, the whole NYT corpus contains a  significant number of them which are mentioned for the first time after the beginning of the document, with \num{44192} salient entities whose first position is after the first 20\% of the news for a total of \num{31128} such news (out of the total \num{108925} news in NYT). On the other hand, the salient entities present in Wikinews are mainly confined at the beginning of documents, with only $28$ salient entities whose first position is after the first 20\% of the news. For this reason, we only consider  NYT as the main testbed for estimating the flexibility of the systems over entities' position, both for its large size and for the wider distribution that salient entities have inside this corpus.
		
		Figure~\ref{fig:f1positions} shows the comparison among the available systems. Performance is computed only over the test set of the NYT, which contains \num{3911} salient entities whose first position is after the first 20\% of the news, with a total of \num{2821} such news (which are \num{9577} in total in the test set). All systems are highly effective on the classification of salient entities mentioned at the beginning of the document, but their behaviour differs significantly when salient entities are mentioned at the documents' end. In this latter case, \textsc{Swat} does not overfit upon the positional feature and, indeed, obtain a high improvement with respect to \textsc{Cmu-Google-ours} which is respectively up to $14\%$ in micro-F1. As a consequence, we can state that \textsc{Swat} is more flexible with respect to salient-entities'  position than \textsc{Cmu-Google}, so that it could be used consistently over other kinds of documents where salient information is not necessarily confined to their beginning.

		\subsection{Error Analysis}
		\label{sub:error-analysis}
		
		In order to gain some insights on \textsc{Swat} performance and possible improvements, we performed here an error analysis focused on two main points.  The first one aims at analyzing how much \textsc{Swat} can drop in performance when a group of features becomes misleading, while in the second one we manually analyze the erroneous predictions made by our entity salience system over a subsample of documents from NYT and Wikinews.

		\medskip\noindent\textbf{Misleading Features.} The first part of the error analysis complement Section~\ref{sec:feature-analysis} with a different analysis of the features: how much a group of features can lower the entity salience performance of our system when the features are ``wrong''?  Accordingly, we decided to collocate our experiment in the extreme of its spectrum: one by one, we force each group of the most relevant features (i.e., position-, \mbox{title-,} annotation-based and relatedness-based features) to  be intentionally misleading (respectively setting a group of features to $0$ or $1$) and then we analyze the micro-F1 performance of \textsc{Swat}, previously trained on the NYT dataset, over the NYT test set.
		
		Forcing a group of features to be set to $0$ or $1$ for an entity $e$ has the following intuitive interpretations. For position-based features, the value $0$ (resp. $1$) means that we are forcing the entity $e$ to appear at the beginning (resp. end) of the document. For title-based features, the value $0$ (resp. $1$) means that we are forcing the entity $e$ to do not appear (resp. to appear) in the title. For frequency-based features, the value $0$ (resp. $1$) means that an entity $e$ never appear (resp. frequently appear) in the document. For annotation-based features, the value $0$ (resp. $1$) means that the entity $e$ has been annotated in the document with low (resp. high) \textsc{Wat} confidence scores. Finally, for relatedness-based features, the value $0$ (resp. $1$) means that the entity $e$ is poorly (resp. highly) related to the other document's entities.
		
		Not surprisingly, forcing features to be expressly misleading addresses \textsc{Swat} to lower its micro-F1 performance of different points, usually in line with the feature analysis we did in  Section~\ref{sec:feature-analysis}. We observed that this experiment presents a common pattern over all features (except for the position-based features): when a group of features is forced to assume the value $0$, \textsc{Swat} becomes more conservative and it generates more false negative errors, whereas when a group of features is forced to assume the value $1$, \textsc{Swat} generates more false positive errors. This means that features with high values make \textsc{Swat} more confident in its predictions, whereas a more conservative attitude is observed when features have very low values. Specifically, the drops in performance for the group of features when set to $0$ (resp. $1$) are: $-1\%$ (resp. $-10.7\%$) for title-based features, $-37.2\%$ (resp. $-6.5\%$) for frequency-based features, $-5.2\%$ (resp.  $-11.6\%$) for annotation-based features and $-2.3\%$ (resp. $-2.8\%$) for relatedness-based features.

		On the other hand, for position-based features we observed the opposite behaviour: when set to the value $0$, \textsc{Swat} is more confident and it generates more false positive errors, whereas when set to the value $1$, \textsc{Swat} become more conservative and it generates more false negative errors. This is not surprising since entities appearing at the beginning of a document are commonly salient. Specifically, the drops in performance for the group of position-based features when setting to $0$ (resp. $1$) are, respectively, $-46\%$ and $-21\%$.
		
		This part of the error analysis is clearly consistent with what we have already found in Section~\ref{sec:feature-analysis}, but it adds several more insights on how much misleading features can wrongly address \textsc{Swat}'s predictions. In particular, non-salient entities appearing at the very beginning of the document can have a dramatic impact  as well as a low frequency for entities that are actually salient. On the other hand, misleading annotation- and title-based features have a medium impact, while misleading relatedness-based features have a minor impact with a drop in performance of only a small margin of points.

		\medskip\noindent\textbf{Manual Inspection.} The second part of the error analysis involves the manual inspection of the predictions of \textsc{Swat} (using all features), performed over a subset of  80 (=40+40) documents from the NYT and Wikinews datasets.  The most significant result we gain is what argued by~\cite{hasan2014automatic}: namely that the deployment of semantic knowledge (i.e., Wikipedia entities) eliminates some errors that originally afflicted keyphrase extraction algorithms. 
		
		However, our error analysis of 80 documents also showed that false-negative errors (i.e., entities classified as non-salient, despite being salient) are mainly due to the position-based features which frequently induce to miss a salient entity because it is not at the beginning of the news. On the other hand, we also noticed that a large percentage of the analyzed news of NYT ($\sim35\%$) and Wikinews ($\sim40\%$) contain false-positive errors which are ground-truth errors: in these cases \textsc{Swat} correctly identifies the salience of an entity, but the ground truth does not label it as salient and so it is unfortunately counted as an error in our tables.
		
		This analysis suggests that  \textsc{Swat} performance could be actually higher than what we claimed before and a better ground-truth dataset should be devised, as we foresee in the concluding section.

			\section{Conclusion and Future Work}
			\label{sec:conclusions}
			
			In this paper, we have studied the problem of entity salience and proposed a novel system, called \textsc{Swat}, that efficaciously identifies the salient Wikipedia entities occurring in an input document. \textsc{Swat} consists of several modules that are able to detect and classify on-the-fly Wikipedia entities as salient or not, based on a large number of syntactic, semantic and latent signals properly extracted via a supervised process which has been trained over millions of examples drawn from the New York Times corpus. The validation process was performed through a large experimental assessment, on which \textsc{Swat} resulted to improve significantly known solutions over all publicly available datasets. We have released  \textsc{Swat} via a Web API that will allow its use in other software tools.

				The lesson learned from the number of experiments and analysis we did clearly concerns the novel set of investigated features: the simple position and frequency features are sufficient for achieving satisfying performance, but only with the deployment of signals coming from the coherence of the annotated entities and their relatedness, an entity salience system can refine its predictions, achieve state-of-the-art performance and improve the quality of its results.

			Our investigation also highlighted three main research directions that we consider worthy of scientific attention. The first one concerns with the improvement of the quality of the NYT dataset (which is the largest one available) by (i) augmenting its annotations with common nouns and (ii) by labeling its ground-truth via a crowdsourcing task rather than a rule-based approach as the one adopted by \cite{dunietz2014new}. The quality of ground-truth is crucial in order to fairly assess the efficacy of the proposed approaches. In our paper we have hypothesized a performance of \swat{} better than the one established in the experiments because of the limitations inherent in the NYT dataset. 
			
			The second research direction clearly concerns the design of more sophisticated techniques to mitigate the erroneous predictions that could be generated from \textsc{Swat} when its features contain misleading values. First, the current confidence scores (i.e, $commonness$ and $\rho$ scores) provided by \textsc{Wat} are based on a simple combination of occurrence mention-entity statistics (computed over Wikipedia) and the relatedness between an annotation and its surrounding annotated entities: a better design of these scores could make \textsc{Swat} more robust and thus mitigate the drops in performance that wrong confidence-scores could provide. Second, we think that design an approach for mitigating wrong position and frequency statistics is yet more necessary and challenging: mitigating their erroneous value seems to be very difficult since we do not have a total control on where an author collocated the salient entities or how much times he decided to mention a salient entity in the text. Starting a news with relevant information and frequently mention it in the text is a standard pattern, but it easily misleads the classifier when a non-salient entity follows it. An approach that can mitigate this problem could be to enhance our system with further knowledge coming from multiple source of news: if entities that are mentioned at the beginning of a document are never mentioned in similar positions in other news that cover the same topics (in the same period of time)  they are probably not salient. Similar considerations can also hold for frequency-based misleading features. 
			
			The third research direction concerns with the design of faster entity linkers which are crucial to allow the processing of large datasets, such as NYT, in a reasonable time. In fact, the current annotation of NYT by \textsc{Wat}, although it runs on multiple threads, took about 20 days. If we wish that academic entity linkers scale to the annotation of Big Data, researchers should concentrate on their engineering by possibly balancing speed with the precision of the annotation. In particular, from our preliminary analysis of \textsc{Wat}, the bottleneck of entity linking systems is caused by the number of candidate entities that is associated to each mention and then used to feed a combinatorial approximation algorithm that outputs the final annotated entities. Despite a consistent number of works is present for the different approximation algorithms that can be used, no prior work (to the best of our knowledge) has yet investigated the design of pruning strategies for proper reducing the candidate entities and then evaluating the trade-off between accuracy and speed that can be achieved from their balancing.

			Finally, we mention the problem of testing our approach and the other known ones over datasets of a different type than news and over other promising domains, such as expert finding in academia \citep{cifariello2019wiser}. Despite a different kind of textual data, our system \textsc{Swat} could actually be able to correctly detect the salient entities both in web pages and research papers in a similar fashion as already done with news: classical information extraction systems \citep{gamon2013identifying,florescu2017positionrank}  designed their algorithms only on the top of positional and frequency signals (which are also deployed by \textsc{Swat}), but without using Wikipedia entities (and relative annotation- and relatedness-based features) as salient elements.

		\vspace{-0.2cm}
		\section*{Acknowledgments} 
		
		We thank the anonymous reviewers for their careful reading of the manuscript and their insightful comments that allowed us to significantly improve the quality of the paper.  Part of the work of the first two authors has been supported by a \textit{Bloomberg Data Science Research Grant (2017)}, and by the EU grant for the Research Infrastructure \textit{``SoBigData: Social Mining \& Big Data Ecosystem''} (INFRAIA-1-2014-2015, agreement \#654024).

		\bibliographystyle{coin}
		\bibliography{main.bib}

		\newpage
		
		\appendix
		
		\section{Graphical User Interface and Public API}
		\label{sub:api}

\begin{figure}[t!]
			\centering
				\includegraphics[width=.8\textwidth]{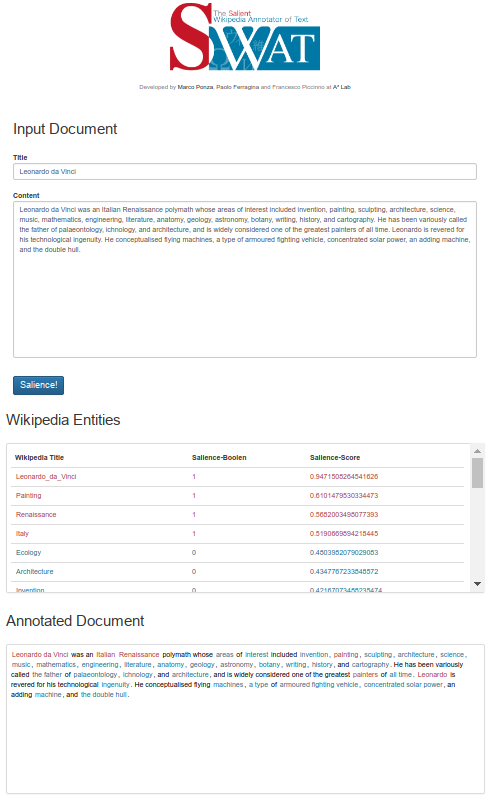}
			\vspace{0.3cm}
			\caption{The GUI of \textsc{Swat} prototype allows detecting and classifying Wikipedia entities from an input text. The box \textsf{Wikipedia Entities} shows the annotated entities with a boolean label, denoting salient (red) and non-salient (blue) entities, and ranked by their \textsf{Salience-Score}, namely {XGboost}'s probability. The box \textsf{Annotated Document} shows the mentions annotated to their pertinent Wikipedia entities.
				\label{fig:swatgui}
			}
		\end{figure}

		Figure~\ref{fig:swatgui} shows a simple GUI\footnote{The demo of the system is accessible at \href{https://swat.d4science.org}{swat.d4science.org}.} that allows using \textsc{Swat} over an input document loaded via a Web interface. In addition to the GUI, it is possible to deploy \textsc{Swat} through a REST-like interface\footnote{The API is accessible at \href{https://sobigdata.d4science.org/web/tagme/swat-api}{sobigdata.d4science.org/web/tagme/swat-api}.}. The API provides results in both human and machine-readable form, by deploying a simple JSON format (see Tables \ref{table:jsonparams-first}, \ref{table:jsonparams-second} and \ref{table:jsonannotations}). In order to show how the interaction with \textsc{Swat} works, we offer a Python code snippet in Listing \ref{lst:pythonswat} for querying our system and the corresponding JSON response in Listing \ref{lst:pythonswat_response}. A query requires just one optional parameter (i.e., title) and one mandatory parameter (i.e., the content of the document). The response includes all entities annotated by \textsc{Swat} and different information for each of them.

		\vspace{0.5cm}
		\begin{python}[basicstyle=\scriptsize\ttfamily, label={lst:pythonswat}, caption={Python code for querying the \textsc{Swat}'s public API. The authorization token \textsf{MY\_GCUBE\_TOKEN} is needed for using the service and obtainable through free registration.}]
import json
import requests

MY_GCUBE_TOKEN = 'copy your gcube-token here!'

document = {
	'title': 'Obama travels.',
	'content': 'Barack Obama was in Pisa for a flying visit.'
}

response = requests.post('https://swat.d4science.org/salience',
                         data=json.dumps(document),
                         params={'gcube-token': MY_GCUBE_TOKEN})

print json.dumps(response.json(), indent=4)
		\end{python}

		
		\vspace{0.5cm}
		
		\begin{python}[basicstyle=\scriptsize\ttfamily, label={lst:pythonswat_response}, caption={Structure of the JSON response of \textsc{Swat}.}]
{
	'status'                         # str
	'annotations':
	{
		'wiki_id'                # int
		'wiki_title'             # str
		'salience_class'         # int
		'salience_score'         # float
		'spans':                 # where the entity is mentioned in content
		[
		    {
		    	'start'          # int (character-offset, included)
		    	'end'            # int (character-offset, not included)
	    	    }
		]
	}
	'title'                          # str
	'content'                        # str
}
			
		\end{python}

		\setlength\extrarowheight{5pt}
		
		\begin{table*}[t!]

			\vspace{2cm}
			
			\caption{Fields of the \textsc{Swat}'s JSON request. }
			
			\label{table:jsonparams-first}

			\resizebox{\textwidth}{!}{
				
				\begin{tabularx}{\textwidth}{l@{\hspace{2cm}}p{9cm}@{\hspace{10pt}}l@{\hspace{10pt}}}
					\toprule
					
					\textbf{Name} & \textbf{Description}   & \textbf{Type} \\ \midrule
					
					\textsf{title}              & Title of the document. & \textsf{String}    \\
					
					\textsf{content}            & Content of the document. & \textsf{String}    \\ \bottomrule 
					
				\end{tabularx}
				
			}

			\vspace{1.5cm}
			
			\caption{Fields of the \textsc{Swat}'s JSON response. }
			\label{table:jsonparams-second}
			
			\resizebox{\textwidth}{!}{
				
				\begin{tabularx}{\textwidth}{l@{\hspace{2cm}}p{8.5cm}@{\hspace{10pt}}l@{\hspace{10pt}}}
					\toprule
					
					\textbf{Name} & \textbf{Description}   & \textbf{Type} \\ \midrule
					
					\textsf{status}             & Status of the response.   & \textsf{String}    \\
					
					\textsf{annotations}   & List of extractions (see Table~\ref{table:jsonannotations}).  & \textsf{List}   \\ \bottomrule 
					
				\end{tabularx}
				
			}

			\vspace{1.5cm}
			\caption{Fields present in each object of \textsf{annotations} field in the JSON response. }
			
			\label{table:jsonannotations}

			\resizebox{\textwidth}{!}{
				
				\begin{tabularx}{\textwidth}{l@{\hspace{10pt}}p{9.3cm}@{\hspace{10pt}}l@{\hspace{10pt}}}
					\toprule
					
					\textbf{Name} & \textbf{Description}   & \textbf{Type} \\  \midrule

					\textsf{wiki\_id} & Wikipedia ID of the extracted entity. & \textsf{Integer} \\ 
					
					\textsf{wiki\_title} & Wikipedia title of the extracted entity. & \textsf{String} \\

					\textsf{salience\_boolean} & $1$ if the entity is salient, $0$ otherwise. & \textsf{Integer} \\

					\textsf{salience\_score} & Score of relevance of the entity. & \textsf{Float} \\

					\textsf{spans} & List of pairs of integers. Each pair contains the start (included) and end (excluded) offsets at character-level of the extracted entity in the input text. & \textsf{List} \\ \bottomrule  
				\end{tabularx}					
			}

		\end{table*}

		\setlength\extrarowheight{0pt}

		\newpage

		\clearpage
		
		
		\label{lastpage}
		
	\end{document}